\documentclass[longauth]{aa} % for the long lists of affiliations 
\usepackage{graphicx}
\usepackage{txfonts}
\begin{document}

   \title{The VIMOS Public Extragalactic Redshift Survey (VIPERS)}

%   \subtitle{The size dependence and evolution of the high-mass limit
%     of the blue cloud and galaxy bimodality over the last eight billion years
\subtitle{Downsizing of the blue cloud and the influence of galaxy size on mass quenching over the last eight billion years
\thanks{based on observations collected at the European Southern Observatory, Cerro Paranal, Chile, using the Very Large Telescope under programs 182.A-0886 and partly 070.A-9007.
Also based on observations obtained with MegaPrime/MegaCam, a joint project of CFHT and CEA/DAPNIA, at the Canada-France-Hawaii Telescope (CFHT), which is operated by the
National Research Council (NRC) of Canada, the Institut National des Sciences de l’Univers of the Centre National de la Recherche Scientifique (CNRS) of France, and the University of Hawaii. This work is based in part on data products produced at TERAPIX and the Canadian Astronomy Data Centre as part of the Canada-France-Hawaii Telescope Legacy Survey, a collaborative project of NRC and CNRS. The VIPERS web site is http://www.vipers.inaf.it/.}}

   \author{C.~P.~Haines\inst{1}
\and A.~Iovino\inst{1}
\and J.~Krywult\inst{2} 
% GROUP A1
\and L.~Guzzo\inst{1,3}      
\and I.~Davidzon\inst{4,5}   
\and M.~Bolzonella\inst{5}      
\and B.~Garilli\inst{6}          
\and M.~Scodeggio\inst{6}       
\and B.~R.~Granett\inst{1}                                                 
\and  S.~de la Torre\inst{4}       
%GROUP A2: 
\and G.~De Lucia\inst{7}
\and U.~Abbas\inst{8}
\and C.~Adami\inst{4}
\and S.~Arnouts\inst{4} 
\and D.~Bottini\inst{6}
\and A.~Cappi\inst{5,10}
\and O.~Cucciati\inst{9,5}           
\and P.~Franzetti\inst{6}   
\and A.~Fritz\inst{6}       
\and A.~Gargiulo\inst{6}
\and V.~Le Brun\inst{4}
\and O.~Le F\`evre\inst{4}
\and D.~Maccagni\inst{6}
\and K.~Ma{\l}ek\inst{11}
\and F.~Marulli\inst{9,12,5} 
\and T.~Moutard\inst{25,4}  
\and M.~Polletta\inst{6,13,14}
\and A.~Pollo\inst{15,11} 
\and L.~A.~M.~Tasca\inst{4}
\and R.~Tojeiro\inst{16}
\and D.~Vergani\inst{17,5}
\and A.~Zanichelli\inst{18}
\and G.~Zamorani\inst{5}
% GROUP B
\and J.~Bel\inst{20}
\and E.~Branchini\inst{21,22,23}
\and J.~Coupon\inst{24}
\and O.~Ilbert\inst{4}
%GROUP C
\and L.~Moscardini\inst{9,12,5}
\and J.~A.~Peacock\inst{26}
\and M.~Siudek\inst{27}
}

   \institute{INAF - Osservatorio Astronomico di Brera, via Brera 28,
     20122 Milano, via. E. Bianchi 46, 23807 Merate, Italy \\ %1
              \email{chris.haines@oa-brera.inaf.it}
\and Institute of Physics, Jan Kochanowski University, ul. Swietokrzyska 15, 25-406 Kielce, Poland %2
\and  Universit\`{a} degli Studi di Milano, via G. Celoria 16, 20133 Milano, Italy %3
\and Aix Marseille Univ, CNRS, LAM, Laboratoire d'Astrophysique de Marseille, Marseille, France  %4
\and INAF - Osservatorio Astronomico di Bologna, via Ranzani 1, I-40127, Bologna, Italy %5
\and INAF - Istituto di Astrofisica Spaziale e Fisica Cosmica Milano, via Bassini 15, 20133 Milano, Italy%6
\and INAF - Osservatorio Astronomico di Trieste, via G. B. Tiepolo 11, 34143 Trieste, Italy %7
\and INAF - Osservatorio Astronomico di Torino, 10025 Pino Torinese, Italy %8
\and Dipartimento di Fisica e Astronomia - Alma Mater Studiorum Universit\`{a} di Bologna, viale Berti Pichat 6/2, I-40127 Bologna, Italy %9
\and Laboratoire Lagrange, UMR7293, Universit\'e de Nice Sophia Antipolis, CNRS, Observatoire de la C\^ote d’Azur, 06300 Nice, France %10
\and National Centre for Nuclear Research, ul. Hoza 69, 00-681 Warszawa, Poland %11
\and INFN, Sezione di Bologna, viale Berti Pichat 6/2, I-40127 Bologna, Italy %12
\and Aix-Marseille Université, Jardin du Pharo, 58 bd Charles Livon, F-13284 Marseille cedex 7, France%13
\and IRAP,  9 av. du colonel Roche, BP 44346, F-31028 Toulouse cedex 4, France%14
\and Astronomical Observatory of the Jagiellonian University, Orla 171, 30-001 Cracow, Poland %15
\and School of Physics and Astronomy, University of St Andrews, St Andrews KY16 9SS, UK %16
\and INAF - Istituto di Astrofisica Spaziale e Fisica Cosmica Bologna, via Gobetti 101, I-40129 Bologna, Italy %17
\and INAF - Istituto di Radioastronomia, via Gobetti 101, I-40129, Bologna, Italy %18
\and Canada-France-Hawaii Telescope, 65--1238 Mamalahoa Highway, Kamuela, HI 96743, USA %19
\and Aix Marseille Univ, Univ Toulon, CNRS, CPT, Marseille, France %20
\and Dipartimento di Matematica e Fisica, Universit\`{a} degli Studi Roma Tre, via della Vasca Navale 84, 00146 Roma, Italy %21
\and INFN, Sezione di Roma Tre, via della Vasca Navale 84, I-00146 Roma, Italy %22
\and INAF - Osservatorio Astronomico di Roma, via Frascati 33, I-00040 Monte Porzio Catone (RM), Italy %23
\and Astronomical Observatory of the University of Geneva, ch. d'Ecogia  16, 1290 Versoix, Switzerland%24
\and Department of Astronomy \& Physics, Saint Mary's University, 923 Robie Street, Halifax, Nova Scotia, B3H 3C3, Canada%25
\and Institute for Astronomy, University of Edinburgh, Royal Observatory, Blackford Hill, Edinburgh EH9 3HJ, UK %26
\and Center for Theoretical Physics, Al. Lotnikow 32/46, 02-668 Warsaw, Poland%27
%              \thanks{The university of heaven temporarily does not
%                     accept e-mails}
             }

\offprints{Chris Haines\\  \email{chris.haines@oa-brera.inaf.it}}
%   \date{Received September 15, 1996; accepted March 16, 1997}

  \abstract{We use the full VIPERS redshift survey in combination with
    SDSS-DR7 to explore the relationships between star-formation
    history (using d4000), stellar mass and galaxy structure, and how
    these relationships have evolved since $z{\sim}1$. 
We trace the extents and evolutions of both the blue cloud and red
sequence, by fitting double Gaussians to the d4000 distribution of galaxies in narrow
stellar mass bins, for four redshift intervals over $0{<}z{<}1$. This reveals
downsizing in star formation, as the high-mass limit of the blue cloud
retreats steadily with time from \smash{$\mathcal{M}\,{\sim} 10^{11.2\,}{\rm
  M}_{\odot}$} at $z{\sim}0.9$ to \smash{$\mathcal{M}\,{\sim}10^{10.7\,}{\rm
  M}_{\odot}$} by the present day. The number density of massive
blue-cloud galaxies (\smash{$\mathcal{M}\,{>}10^{11\,}{\rm M}_{\odot}$},
d$4000{<}1.55$) drops sharply by a factor five between $z{\sim}0.8$
and $z{\sim}0.5$. These galaxies are becoming quiescent at a rate that
largely matches the increase in the numbers of massive passive
galaxies seen over this period.
We examine the size-mass relation of blue cloud
 galaxies, finding that its high-mass boundary runs along lines of
    constant \smash{$\mathcal{M}/r_{e}$} or equivalently inferred velocity
    dispersion. Larger galaxies can continue to form stars to higher
    stellar masses than smaller galaxies. 
As blue cloud galaxies approach this high-mass limit,
    they start to be quenched, their d4000 values increasing to
    push them towards the green valley. In parallel, their structures change, showing higher S\'{e}rsic
    indices and central stellar mass densities. 
For these galaxies, bulge growth is necessary for them to reach the high-mass limit of
    the blue cloud and be quenched by internal mechanisms. 
The blue cloud galaxies that are being quenched at $z{\sim}0.8$ lie along the
same size-mass relation as present day quiescent galaxies, and seem the likely progenitors of today's S0s.}


   \maketitle
%
%-------------------------------------------------------------------

\section{Introduction}

The recent wide-field surveys such as the Sloan Digital Sky Survey (SDSS) have firmly established that galaxies in the local Universe can be
broadly divided into two distinct populations according to their
UV-optical colours: the blue cloud made up of young, star-forming galaxies; and the 
red sequence of passively-evolving galaxies that becomes
increasingly dominant with stellar mass \citep{strateva,baldry04,wyder,blanton09,taylor}. 
This bimodality in colour (or star-formation activity) and its
mass-dependence has been shown to persist to at least $z{\sim}4$ \citep{pozzetti,muzzin,ilbert13,tomczak}, with
star-forming galaxies following a tight relation between star
formation rate (SFR) and stellar mass ($\mathcal{M}$) \citep{noeske,speagle,lee}.

This dichotomy in colour is deeply entwined with the well-known
structural bimodality of galaxies, which can broadly be separated into 
smooth early-types and disc-dominated spirals
\citep{hubble,sandage,kormendy}. The bivariate distribution of galaxies in
the colour--S\'{e}rsic index ($\eta$) plane reveals two distinct
peaks corresponding to blue, star-forming discs and red, quiescent
bulges \citep{driver06,ball}, up to $z{\sim}2$ \citep{bruce,krywult}.
The star-formation activity of galaxies in the local Universe has been
found to correlate more strongly with mean stellar mass density
\citep[$\Sigma_{e}$;][]{k03,k06,brinchmann}, the presence of a bulge
\citep{bell08}, and the stellar mass density
within the central kpc \citep[$\Sigma_{1}$;][]{fang} than with stellar
mass. \citet{omand} found that the fraction of quiescent galaxies
correlates best with $\mathcal{M}/r_{e}^{3/2}$. These correlations
have been found to persist up to $z{\sim}3$ \citep{franx,wuyts,cheung,bell12,lang,barro,whitaker16}. 
These correlations suggest that the internal structure of a galaxy,
and its development, must play a crucial role in driving or regulating its star
formation activity, although \citet{lilly} show how these correlations
can arise naturally even in cases where structure plays no physical role.

The total stellar mass content of galaxies has approximately
doubled since $z{\sim}1$.0--1.2 \citep[e.g.][]{muzzin,ilbert13,tomczak}. This growth in the
integrated stellar mass contained within galaxies has been found to be
in good agreement with the cosmic SFR integrated over the same period \citep{bell07,ilbert13}.
However, while these new stars are being
formed within the blue cloud, the build-up of stellar mass
is largely confined to the red sequence population of quiescent
galaxies \citep{arnouts07,tomczak}, while the integrated stellar mass within the star-forming galaxy population has remained
approximately constant since $z{\sim}1$ \citep{bell07,ilbert10,pozzetti}. 

If the massive blue-cloud galaxies observed at $z{\sim}1$ were to
continue growing by forming stars at rates given by the evolving 
 SFR--$\mathcal{M}$ relation of normal star-forming galaxies through
 to the present day, this would leave a large population of massive
${\ga}10^{11}{\rm M}_{\odot}$ spiral galaxies at $z{\sim}0$ \citep{bell07}.
This population is simply not seen in the local universe. 
Instead, the maximum stellar mass of actively star-forming galaxies
was observed by \citet{cowie} to have fallen steadily from $z{\sim}1$
to lower redshifts, a process they termed as ``downsizing''.
The term downsizing has since been used to describe a variety of
seemingly anti-hierarchical behaviours within galaxy populatons \citep{fontanot}, such as the
finding that massive quiescent galaxies have stellar populations
that formed earlier and over shorter time-scales than lower-mass systems
\citep[e.g.][]{thomas}. This can lead to confusion. Here we refer
to downsizing in the original sense of \citet{cowie}.

Something is required to shut down star formation in massive blue cloud
galaxies at a high enough rate to ensure that the stellar mass function (SMF) of star-forming
galaxies remains approximately unchanged since $z{\sim}1$
\citep{pozzetti,moutard}, and to enable the high-mass end of the red sequence to build up rapidly
over the same period \citep{davidzon13,fritz}. \citet{peng10} show that to keep the blue cloud
SMF unchanged requires a mechanism which terminates
star formation in massive galaxies at a rate that is statistically
proportional to their SFR. This process is loosely termed ``mass quenching''
and is believed to be the dominant mechanism for terminating star
formation among high-mass (${\ga}10^{10.5}{\rm M}_{\odot}$) blue cloud
galaxies, while environment-related processes become more important at
lower stellar masses \citep[e.g.][]{haines07}. In this work,
we focus on those quenching processes that act internally to the
galaxy (including its circumgalactic medium) and which define the limiting
galaxy properties beyond which they cannot continue growing through forming stars. We do not consider environmental effects further.

A variety of physical mechanisms have been proposed that could be
responsible for quenching star formation in blue cloud galaxies once
they reach a certain mass \citep{gabor}. Feedback from an active galactic nucleus (AGN)
has been frequently invoked to limit the growth of massive galaxies and
produce the turnover in the SMF by suppressing
gas cooling \citep{bower06}. Gas-rich galaxy mergers channel
large amounts of gas onto the central nucleus, fuelling powerful
starbursts and rapid black hole growth, until feedback from accretion
is able to drive quasar winds and expel the remaining gas from the
galaxy, quenching star formation \citep{dimatteo,hopkins}. Subsequent low-level ``radio mode'' AGN feedback
may then prevent the hot gas corona from cooling and reinitiating star
formation \citep{croton}.

\citet{martig} proposed that the build up of a central mass
concentration or bulge within a massive star-forming galaxy could
stabilize the gas disc against fragmentation and collapse into
molecular gas clouds, and morphologically quench its star
formation from the inside-out \citep{genzel}. Low-level feedback from
the winds of AGB stars could contribute to prevent the residual gas
from cooling in dense, early-type galaxies, keeping them quiescent
\citep{conroy}. 
The evolution of a galaxy is also thought to depend on the mass of the
dark matter (DM) halo that hosts it, through the transition from
low-mass halos where narrow streams of cold gas continually feed the
galaxy with new fuel for star formation, to DM haloes more massive than 
${\sim}10^{12\,}{\rm M}_{\odot}$ where stable virial shocks form,
heating the infalling gas to the virial temperature (${\sim}10^{6}$K),
and dramatically reducing the accretion rate of gas onto the galaxy \citep{keres,dekel06,dekel09}

The star formation history of a galaxy is fundamentally written within
its spectrum, comprising both emission lines from gas ionized by the
UV photons of short-lived hot O and B stars, representing
the current SFR, and the stellar continuum
produced by evolved stars. This continuum contains a complex array of spectral features and
absorption lines that encode many details of the stellar demographics.
One of the most direct methods for characterising the star formation
history of galaxies is measuring the 
4000\smash{\AA} spectral break (hereafter d4000), the strongest
discontinuity in the optical spectrum of a galaxy. The strong break
occurs in older stellar populations due to a sudden accumulation of
absorption features blueward of 4000\smash{\AA} (e.g. Ca{\sc ii} H+K lines) that appear in stars cooler
than G0 (6000K) due to ionized metals. In hotter stars, elements are multiply ionized and
the opacity decreases, so the break appears much smaller (d$4000{\simeq}1$.0--1.4). 
Since hot stars are short-lived, the d4000 index increases
monotonically with stellar age. 
\citet{k03b} showed that it is an excellent age indicator, in
particular for young (${<}1$\,Gyr, d$4000{<}1.5$) stellar
populations where it does not depend strongly on metallicity. 
\citet{brinchmann} showed that d4000 may also be used as a proxy for
specific-SFR (SFR/$\mathcal{M}$). 
\citet{k03} showed that galaxies in the local Universe can be split
into two well-separated populations: low-mass galaxies with low d4000
values (${<}1.55$) indicative of young stellar populations, low
surface mass densities and S\'{e}rsic indices typical of discs; and
high-mass galaxies with high d4000 values (${>}1.55$) indicative of
old stellar populations, high stellar mass densities and
concentrations. Later studies that combined d4000 with a range
of age and metallicity-dependent spectral indices to obtain more robust
constraints on the light-weighted stellar ages of galaxies, confirmed that the
bimodality in d4000 seen by \citet{k03} indeed signifies a bimodality
in stellar age \citep{gallazzi,haines06}.

The key limitation of \citet{k03} and other SDSS-based studies is that
all the galaxies are at essentially the same redshift, so that we
cannot follow the evolution of the observed trends between d4000 and
$\mathcal{M}$ or galaxy structure over a cosmically signficant time-scale, or determine when these trends were
established. The present paper attempts to do this, taking advantage of
the recently completed VIMOS Public Extragalactic Redshift Survey
(VIPERS) to trace the bimodal distribution of galaxies in the
d4000--$\mathcal{M}$ plane back to $z{\sim}1$, and explore how the
relationships between star-formation history, stellar mass and galaxy
structure have evolved over the last eight billion years.

The fundamental objective of the VIPERS project has been to provide a
representative vision of the large-scale structure and the global
properties of galaxies when the Universe was about half its current
age \citep{guzzo}, just as the SDSS survey has done for the local Universe. 
The availability of robust spectral parameters including d4000 and [O{\sc
  ii}], as well as stellar masses and structural parameters obtained
from high-quality CFHTLS imaging, for a well-defined sample of almost
100,000 galaxies at $0.5{<}z{<}1.2$, allows the distributions of the
global properties of galaxies to be mapped in fine detail. 
VIPERS uniquely provides a large volume coverage at $0.5{<}z{<}1.2$
allowing the assembly of statistically representative samples of
relatively rare massive galaxies above $10^{11\,}{\rm M}_{\odot}$, and in
  particular those at the high-mass end of the blue cloud. The VIPERS
  survey and the local SDSS sample are described in
  Section~\ref{sec_data}.

Section~\ref{sec_bimodal} examines the evolution of the bimodal
distribution of galaxies in the d4000--\smash{$\mathcal{M}$} plane, and
downsizing in the high-mass limit of the blue cloud. 
The evolution in the number density of massive blue-cloud galaxies is
presented in Section~\ref{sec_m11}. 
Section~\ref{sec_sizemass} presents the size-mass
relations of blue-cloud galaxies, examining the dependencies of d4000
and $\eta$ on the effective radius and stellar mass of galaxies, and
demonstating how mass quenching is influenced also by galaxy size.
The discussion and summary follow in Sections~\ref{sec_discussion} and~\ref{sec_summary}.
A concordance $\Lambda$CDM cosmology with $\Omega_{M}{=}0.3$,
$\Omega_{\Lambda}{=}0.7$ and $H_{0}{=}70$\,km\,s$^{-1}$\,Mpc$^{-1}$ is
assumed throughout.

\section{Data}
\label{sec_data}

VIPERS is a spectroscopic survey, completed in 2015, which has
targeted ${\sim}100$\,000 galaxies at $0.5{\la}z{\la}1.2$ over 23.5\,deg$^{2}$, split
into two fields within the Canada-France-Hawaii Telescope Legacy
Survey Wide (CFHTLS-Wide\footnote{http://www.cfht.hawaii.edu/Science/CFHTLS/}), namely W1 and W4. All details of the survey
design, construction and scientific goals can be found in \citet{guzzo}. 
Spectroscopic targets were selected to have $17.5{\le}i_{AB}{\le}22.5$
(after correction for Galactic extinction), with a second selection
criterion, based on ($g-r$) and ($r-i$) colours applied to exclude
low-redshift ($z{<}0.5$) objects. 

All the spectroscopic observations for VIPERS were carried out using the
VIsible Multi-Object Spectrograph (VIMOS) on the 8.2m Very Large
Telescope (VLT) Unit 3, with the LR-Red
grism which provides a spectral resolution $R{\sim}220$ over the
wavelength range 5500--9500\smash{\AA} at a dispersion of 7.3\smash{\AA}/pixel.
To maximize the multiplex capabiliity of VIMOS, the short-slit
strategy described in \citet{scodeggio09} was used. This enabled a
target sampling rate of 47\% to be reached with a single pass. 
The reduction and redshift measurement is performed within a fully
automated pipeline, before each redshift is independently validated by
two team members and its reliability quantified using the flag
$z_{\rm flag}$, following a scheme similar to that used for VVDS
\citep{lefevre} and zCOSMOS \citep{lilly07}. From repeated
measurements, typical redshift uncertaintes of
$\sigma_{z}{=}0.00054(1+z)$ are obtained \citep{scodeggio16}.

The dataset used in this paper is based on an internal pre-release of the final,
complete VIPERS spectroscopic catalogue, the Second Public Data
Release \citep[PDR-2;][]{scodeggio16}. From this sample, we consider only the 75\,479 galaxies with highly accurate redshift measurements, that
is with $2{\le}z_{\rm flag}{\le}9$ and $z{>}0$. This excludes broad-line AGN, stars and
secondary objects that happened by chance to appear in the slit of a primary target. 
The strength of the 4000\smash{\AA} break
was measured for all galaxies with reliable redshifts in the range $0.414{<}z{<}1.346$, which
meant that the spectral feature was fully contained within the wavelength range
covered by the VIMOS spectra \citep{garilli}. We adopt the narrow definition of the
amplitude of the 4000\smash{\AA} break introduced by \citet{balogh99}, which
is relatively insensitive to the effects of dust reddening. This is
defined as the ratio of the average continuum flux density $F_{\nu}$
in the wavebands 4000--4100\smash{\AA} and 3850--3950\smash{\AA}.
At $z{<}0.5$ the VIPERS sample becomes highly incomplete due to the
$ugri$ colour cuts imposed in the selection of spectroscopic targets,
while at $z{>}1.1$, the survey becomes incomplete even at the
highest stellar masses \citep[\smash{$\mathcal{M}{>}10^{11\,}{\rm M}_{\odot}$};][]{davidzon13}.
We thus limit our analysis to redshifts $0.5{\le}z{<}1.1$, giving
a final sample of 65\,142 galaxies with reliable redshifts and d4000
measurements, 43\,607 in the W1 field and 21\,535 in W4.
The PDR-2 sample includes 450 additional reliable redshifts in the
range $0.5{\le}z{<}1.1$ that were validated after the analysis was
completed, i.e. a difference of ${<}1$\% from the sample used.

Since it was only possible to obtain redshifts for ${\sim}4$0\% of all
possible targets, statistical weights are required to make this
subsample representative of all $0.5{\le}z{<}1.1$ galaxies within the
parent photometric catalogue ($i_{AB}{\le}22.5$).
For every galaxy in the final sample, the likelihood that it was targeted
and a reliable redshift obtained is encoded through the Target
Sampling Rate (TSR) and Spectroscopic Success Rate (SSR)
parameters \citep[for full details see][]{guzzo}. The TSR reflects the localized variation in the spatial
density of targets across the survey area, as the fraction of galaxies
within a given VIMOS quadrant satisfying the selection criteria that
it was then possible to place behind a slit and observe. The SSR
represents the probability of being able to obtain a reliable redshift
for a galaxy, given its $i_{AB}$ magnitude, redshift and VIMOS
quadrant. 

For each galaxy in the VIPERS spectroscopic sample, improved
photometric measurements have been obtained in the $ugriz$ bands from
the T0007 release of the CFHTLS images and photometric catalogues of
the W1 and W4 fields, combined with $K_{s}$ photometry from the VIPERS Multi-Lambda Survey
\citep[VIPERS-MLS\footnote{http://cesam.lam.fr/vipers-mls/};][]{moutard1},
based on follow-up CFHT/WIRCam $K_{s}$-band imaging or from the VISTA
Deep Extragalactic Observations \citep[VIDEO;][]{jarvis} survey.
Total stellar masses (\smash{$\mathcal{M}$}) and rest-frame colours are
obtained using the updated {\em Hyperzmass}
\citep{bolzonella,bolzonella10} code, with a set-up similar to that used in \citet{davidzon16}.
The $ugrizK_{s}$ photometry of each galaxy is
compared to a set of synthetic spectral energy distributions (SEDs)
shifted by $1{+}z_{\rm spec}$. The SED library is built using the stellar
population synthesis model of \citet{bruzual}, with a wide variety of
star-formation histories and adopting the universal
initial mass function (IMF) of \citet{chabrier}. Each template has a
fixed metallicity ($Z{=}0.004$ or $Z{=}0.02$) and dust
reddening that follows \citet{calzetti} or Pr\'{e}vot-Bouchet
\citep{prevot,bouchet} attenuation curves ($A_{V}$ ranging from 0 to
3). The code selects as best-fit SED the model that minimizes the
\smash{$\chi^{2}$} value. The stellar mass estimate is taken
directly from the best-fit template. Absolute magnitudes in the
rest-frame waveband $\lambda_{0}$ are computed starting from the apparent magnitude
observed in the filter closest to \smash{$\lambda_{0}/(1+z)$}, and then
$k$-corrected using the model SED.
We derive the stellar mass completeness limits of the VIPERS survey as
a function of both redshift and d4000 value as described in Appendix~\ref{smcomplete}.

\subsection{Structural parameters}

\citet{krywult} determined robust structural parameter measurements
(e.g. $r_{e}$, $\eta$) for
VIPERS galaxies with reliable redshifts ($2{\le}z_{\rm flag}{\le}9$), fitting
2D PSF-convolved S\'{e}rsic profiles to the observed $i$-band light 
distribution of galaxies from the high-quality CFHTLS-Wide images (T0006 release). 
These $1^{\circ}{\times}1^{\circ}$ images obtained with the MegaCam
instrument have a 0.186$^{\prime\prime}$ pixel scale and typical seeing of
0.64$^{\prime\prime}$ FHWM. To accurately model the PSF variation
across each $1^{\circ}{\times}1^{\circ}$ MegaCam image, elliptical
Moffat fits were performed for 
${\sim}2000$ isolated, bright, non-saturated stars that are distributed across
the full image. Each
fitted parameter of the Moffat function was then approximated by a 2D
Chebyshev polynomial, enabling the local PSF to be generated at the position
of each galaxy. 

In this work we use the circularized $i$-band half-light radius
($r_{e}$) as a measure of each galaxy's size, and the best-fit
S\'{e}rsic index $\eta$ as a measure of its structure. 
\citet{krywult} fully tested the reliability of their
structural parameter measurements by placing model galaxies within the
CFHTLS images, from which they estimate uncertainties in their half-light
radii of 4.4\% (12\%) for 68\% (95\%) of galaxies. 
 Assuming that the stellar mass is distributed radially following the
observed $i$-band light (i.e. there are no radial variations in the
mass-to-light ratio within the galaxy), the effective stellar mass surface density is then defined as
\smash{$\Sigma_{e}{=}\mathcal{M}/(2\pi r_{e}^{2})$}. 
The central stellar mass surface density $\Sigma_{1}$ is estimated by integrating
the best-fit S\'{e}rsic profile in two dimensions out to a radius of
1\,kpc. 

Approximately 13\% of the VIPERS survey area was not covered by the
structural parameter analysis of \citet{krywult}, while other galaxies
are excluded due to having poor local PSFs, GALFIT not converging,
or because the best-fit profile had $\eta{<}0.2$ which is not physical. 
This results in a sample of 49\,533 galaxies with reliable
structural parameters (and reliable redshifts) or 76\% of our full sample of 65\,142. 
The likelihood that we were able to determine structural parameters
for a galaxy shows no sign of depending on its properties
(e.g. stellar mass) or redshift. The stellar mass and redshift distributions of those galaxies with
reliable structural parameters are indistinguishable from the full sample.

\subsection{SDSS comparison sample of local galaxies}

For a comparison sample of local ($z\,{\la}\,0.1$) galaxies with available
stellar masses, d4000 measurements and structural parameters we take the catalogue
of \citet{omand} which is based upon the 7th data release of the SDSS
\citep[SDSS-DR7;][]{sdssdr7}, and matches the structural parameter measurements of
\citet{simard} with the stellar mass estimates and spectral index
measurements (e.g. d4000) from the MPA-JHA DR7 catalogues
\citep{k03b,brinchmann,salim}. 
We consider only those galaxies within the contiguous region of the
the North Galactic Cap, which covers ${\sim}$7\,500\,deg$^{2}$. 
These stellar masses assume a Kroupa IMF, and so we reduce them by
10\% to correct their mass-to-light ratios to the \citet{chabrier} IMF
used here.

For direct comparison to our structural parameters, we take the
$r$-band half-light circularized effective radii (\smash{$r_{e}$}) and S\'{e}rsic index
$\eta$ from the best-fit PSF-convolved pure S\'{e}rsic profiles
obtained by \citet{simard} using GIM2D. \citet{simard} also performed
full bulge-disc decompositions, with $\eta{=}4$ bulge and
exponential disc components, from which we take their measurements of
the fraction of $r$-band light in the bulge component (B/T). The faint galaxy magnitude
limit of the SDSS spectroscopic survey is \smash{$r_{\rm Petro}{=}17.77$}, while
\citet{simard} apply a bright galaxy cut at \smash{$r_{\rm Petro}{=}14.0$}.
This results in incompleteness, both at low masses (\smash{${\la}\,10^{10\,}{\rm
  M}_{\odot}$}) for $z\,{\ga\,}0.055$, but also at the very highest
stellar masses (\smash{${\ga}\,10^{11\,}{\rm M}_{\odot}$}) within $z{\sim}0.04$.

\section{The bimodal d4000 distribution of galaxies and its evolution
  from z=1.0 to the present day}
\label{sec_bimodal}

\begin{figure*}
  \centering
  \includegraphics[width=15cm]{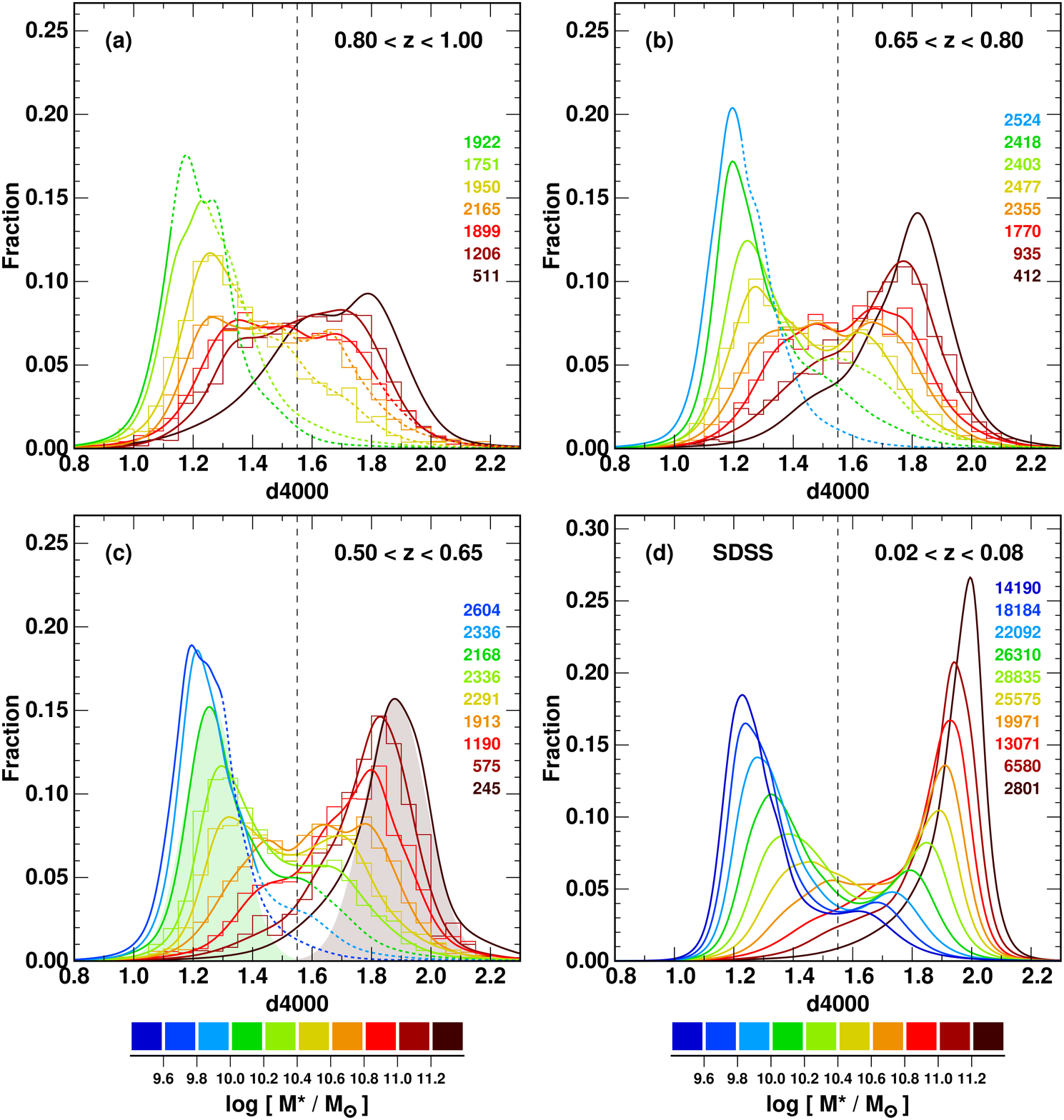}
  \caption{The bimodal d4000 distribution of galaxies as a
    function of stellar mass and redshift. Panels a--c show the d4000 distribution of VIPERS galaxies in
    0.2\,dex wide bins of stellar mass (coloured curves), colour coded
    as indicated, for three redshift bins: 0.8--1.0, 0.65--0.80 and
    0.50--0.65. Panel d does the same, but for galaxies at
    $0.02{\le}z{<}0.08$ taken from SDSS-DR7. 
The curves become dotted in regions of parameter space
   below the stellar mass completeness limits of the VIPERS survey,
   shown as the dashed curves in Fig.~\ref{d4000_smass_evol}. 
For some stellar mass bins, the d4000 distributions are
also shown via histograms with bins of
width 0.05 in d4000. The y-axis scale indicates the fraction of galaxies
within these bins of width 0.05 in d4000. 
   The coloured numbers down the right-hand side
   indicate the number of galaxies in each stellar mass bin. The vertical dashed line indicates the
    ${\rm d4000}{=}1.55$ limit used to separate the blue cloud and red sequence
    galaxy populations. Example Gaussian fits to the d4000
    distributions of blue cloud and red sequence populations in a
    single mass bin are shown by the solid coloured areas.}
  \label{d4000_dist}
\end{figure*} 

\begin{figure*}
  \centering
  \includegraphics[width=18.4cm]{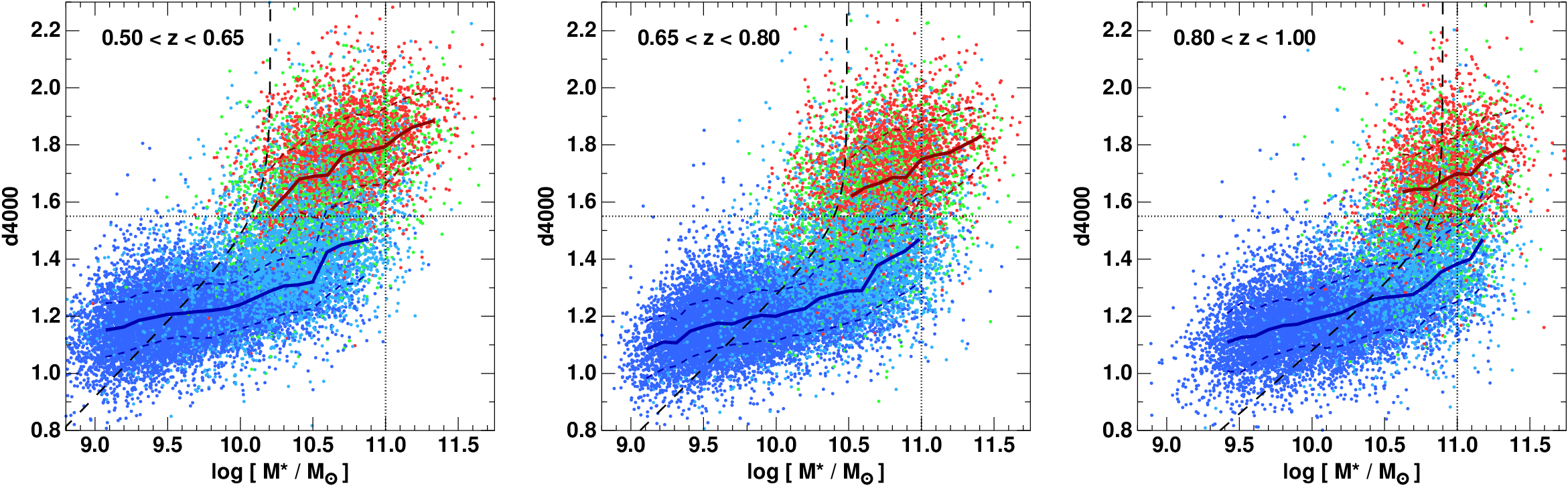}
  \caption{The d4000 versus stellar mass relation of VIPERS galaxies
    in three redshift bins: 0.50--0.65, 0.65--0.80 and
    0.80--1.00. Galaxies are colour coded according to their location
    in the NUV$rK$ diagram of \citet{moutard}: red points
    indicate passive galaxies, green points mark those in the
    green valley, and blue points for those classified as
    star-forming. Dark blue points mark galaxies with [O{\sc
      ii}] emission detected at ${>}3{\sigma}$. Solid red and blue
    curves show the central d4000 values of the red sequence and blue
    cloud populations as a function of stellar mass, while the
    dashed curves mark the $1{\sigma}$ widths of each sequence.
Black dashed curves
    mark the stellar mass completeness limit as a function of
    d4000 for galaxies at the upper redshift limit for each
    panel. Horizontal and vertical dotted lines indicate the
    d$4000{=}1.55$ divider and $\log\mathcal{M}{=}11.0$ mass limit.}
 \label{d4000_smass_evol}
\end{figure*}

Figure~\ref{d4000_dist} shows the d4000 distributions of galaxies,
sliced into 0.2\,dex wide bins of stellar mass, in four redshift
ranges from $z{=}1$ to the present day, combining VIPERS and
SDSS datasets. The VIPERS survey is split into three redshift ranges 
0.8--1.0, 0.65--0.80 and 0.50--0.65 (panels a--c), that span
approximately equal periods of time  (0.89, 0.81 and
0.97\,Gyr respectively). To allow direct comparison with galaxies in the
local Universe, Fig.~\ref{d4000_dist}d presents the d4000
distributions of $0.02{\le}z{<}0.08$ galaxies from SDSS-DR7 divided into
the same stellar mass bins. The mean redshifts of these four bins are
$\langle z\rangle{=}0$.90, 0.73, 0.58 and 0.06 respectively.

The
d4000 distributions, \smash{$f(x)$}, for each stellar mass bin (coloured
curves) are estimated using the
adaptive kernel estimator \citep{silverman}, with each galaxy represented by a Gaussian
kernel whose width $\sigma_{i}$ is proportional to \smash{$f(x_{i})^{-1/2}$}, where $x_{i}$
is the d4000 value of that galaxy. 
By matching the level of smoothing to the density of
information, the adaptive kernel gives an objective and
non-parametric estimator of the underlying distribution of a set of
points, optimised to identify sub-structure and bimodality \citep{pisani}. 

For the first three redshift intervals, every VIPERS galaxy is weighted by the TSR and SSR to reflect the probability
it was targetted and a reliable redshift obtained, as well as an
additional correction to account for stellar mass incompleteness.
To account for stellar mass incompleteness in the SDSS sample, each
galaxy in Fig.~\ref{d4000_dist}d is weighted according to the fraction
of the total volume within $0.02{\le}z{<}0.08$ where it would have
$14.0{\le}r_{\rm Petro}{<}17.77$.
The area under each curve is normalized to unity, so the curves
describe the probability distribution of d4000 within that stellar
mass bin. The curves become dotted when ${>}50$\% of the galaxies at a given d4000, stellar mass and
redshift bin are expected to fall below our $i_{AB}{=}22.5$\,mag survey limit.

Figure~\ref{d4000_dist} shows the development of bimodality in the d4000
distribution over the last 8 billion years. At $0.8{\le}z{<}1.0$
(Fig.~\ref{d4000_dist}a) there
is a robust peak at d$4000\,{\simeq}1.2$ marking the blue cloud, but
only a weak possible peak at d$4000\,{\simeq}1.8$ to mark the red sequence in the highest stellar
mass bin. Otherwise the red sequence appears more as a flat shelf in
the subsequent stellar mass bins (orange/red curves). 
Less than 1\,Gyr later at $0.65{\le}z{<}0.8$ (Fig.~\ref{d4000_dist}b),
the bimodality is now established with a clear peak at
d$4000\,{\simeq}1.8$ in the highest two stellar mass bins 
(\smash{$\log_{10}\!\mathcal{M}\,{\ge}11.0$}). The height of the peak for the red sequence
continues to rise with time through Figs~\ref{d4000_dist}c,d, as does the number of stellar mass bins
where this peak is seen. The bimodality gets continually stronger 
with time, while the valley between the two peaks gets steadily deeper
and wider to the present day. This bimodality allows us to robustly
separate galaxies into blue cloud and red sequence populations in the
VIPERS data by splitting the sample at a d4000 value of 1.55 (dashed line), midway between
the two peaks at $0.50{\le}z{<}0.65$ (Fig.~\ref{d4000_dist}c). This
appears a reasonable choice also at higher redshifts
(Fig.~\ref{d4000_dist}a,b), and is the same limit used by \citet{k03}
to separate young and old galaxies in the SDSS survey.
At all redshifts, the relative fraction of galaxies in the blue cloud decreases with increasing stellar mass, while that in the
red sequence increases. Both blue cloud and red sequence populations systematically shift to
larger d4000 values with increasing stellar mass.

At low to intermediate stellar masses (\smash{$\mathcal{M}\,{\le}10^{10.6\,}{\rm M}_{\odot}$}), the d4000 distribution of blue
cloud galaxies can be well described by a Gaussian function in all
four redshift bins. Where the red sequence shows a strong peak, it
also has a Gaussian-like form. 
Following \citet{baldry} we fit double Gaussians to the d4000
distributions of each stellar mass bin, allowing us to derive the
central d4000 values and dispersions for both blue cloud and red
sequence populations. 

At higher stellar masses ($10.6\,{\la}\log_{10}\!\mathcal{M}\,{\la}11.0$), the blue cloud becomes comparable in
strength or sub-dominant to the red sequence, appearing as either a
small hump or a flat shelf in the d4000 distribution at d$4000\,{\simeq}1$.4--1.5.
This feature is sufficiently prominent to allow the central
d4000 value of the blue cloud to be fitted, and the low-d4000 side modelled by a Gaussian function,
enabling its standard deviation to be estimated. At these stellar
masses there is no obvious green valley: the d4000 distribution
remains flat or rising through to the red sequence, suggestive of a
continuous flow of {\em quenching} galaxies leaving the blue cloud and
moving across to build up the red sequence. 
The typical uncertainties in the d4000 values of galaxies
at these stellar mass ranges are ${\sim}0.05$ (0.07) for those in the blue
cloud in the $0.50{\le}z{<}0.65$ ($0.80{\le}z{<}1.0$) redshift bin,
rising slightly to ${\sim}0.08$ (0.10) in the red sequence, and hence
are not sufficiently high to smooth away any green valley.

In the very highest stellar mass bin(s) ($\log_{10}\!\mathcal{M}{>}11.0$),
the blue cloud appears no more than an extended low-d4000 tail of the
dominant red sequence, and there is no feature enabling the central
d4000 value of the blue cloud to be constrained. The blue cloud galaxies appear too rare to allow us to define
the d4000 distribution of the blue cloud population in any meaningful
way. 
The upper limit of the blue cloud in each redshift slice is determined
as the highest stellar mass bin where a Gaussian function can be
fitted to the d4000 distribution at  d$4000{<}1.55$.

Figure~\ref{d4000_smass_evol} shows the distribution of galaxies in
the \mbox{d4000--$\mathcal{M}$} plane for the same three redshift ranges. 
Solid curves show the central d4000 values of the blue and
red sequences obtained from the double Gaussian fits as a function of mass, while the dashed lines indicate
the 1$\sigma$ widths of each sequence. 
Galaxies are colour coded according to their location in the NUV$rK$
(${\rm NUV}-r$ versus $r-K$) colour-colour diagram, applying the same
delimiting cuts as \citet{moutard} to split the NUV$rK$ diagram
into three regions, enabling galaxies to be classified into passive
(red points in Fig.~\ref{d4000_smass_evol}),
star-forming (blue points) and intermediate ``green valley'' (green
points) populations. The advantage of the NUV$rK$ diagnostic over
traditional single-colour classifications (e.g. $U-V$) is its ability
to effectively disentangle the effects of dust extinction and
star-formation activity on galaxy colours \citep{arnouts}. This
resolves the issue of dusty-red star-burst galaxies contaminating the
red sequence passive population which plagues simpler ($U-V$)-type
classifiers \citep[e.g.][]{haines08,brammer11}.

Figure~\ref{d4000_smass_evol} demonstrates the good consistency 
between the classification of galaxies into blue cloud and red
sequence populations from their d4000 value and that using the NUV$rK$
diagnostic, at all stellar masses and redshifts covered by VIPERS. In the
$0.50{<}z{<}0.65$ bin, 89\% of NUV$rK$-selected passive
galaxies have d$4000{>}1.55$,  
placing them along the red sequence in the d4000 vs stellar mass
diagram, while 92\% of NUV$rK$-selected star-forming galaxies have
d$4000{<}1.55$ and lie along our blue sequence. Conversely, 95\% of
blue cloud galaxies with d$4000{<}1.55$ would be classified as
star-forming from their NUV$rK$ colours. NUV$rK$-selected green valley
galaxies share a similar distribution in the d4000--$\mathcal{M}$ plane to
passive ones, and 72\% have d$4000{>}1.55$. 

An advantage of d4000 over the NUV$rK$ diagnostic is that is provides
a single continuous variable which can be readily associated to the
mean stellar age or star-formation history of a galaxy \citep{k03} 
\citep[although see][for how NUV$rK$ colours can be used to
estimate SFRs]{arnouts}.
This enables us to examine the {\em distribution} of
d4000 values (or equivalently stellar ages), rather than simply split
galaxies into two or three sub-populations. We seek to identify subtle
changes in the d4000 values within sub-populations. These could be indicative of
ongoing changes within these populations due to the slow
decline of star-formation activity in normal blue-cloud galaxies at these redshifts, but also the
first stages of quenching that take a galaxy from the blue cloud to
the red sequence. It also allows the properties
of VIPERS galaxies at $0.5{<}z{<}1.0$ to be directly compared with those in the local Universe
covered by the SDSS \citep{k03,k06}.

\subsection{Evolution of the blue cloud since $z{\sim}1$} 
\label{sec_bluecloud}

Figure~\ref{d4000_dist}  allows the evolution
of the blue cloud population to be followed in detail from $z{=}1$ to the
present day by carefully comparing the d4000 distributions at fixed
stellar mass through the four redshift bins studied.

At $0.8{\le}\,z{<}1.0$ (Fig~\ref{d4000_dist}a), the 
\smash{$11.0{\le}\log_{10}\!\mathcal{M}{<}11.2$} stellar mass bin is the highest one in which the
blue cloud can be defined, as a flat shelf that extends over
d$4000\,{\simeq}1$.3--1.6 containing a
  significant fraction of the galaxy population in that mass bin. 
This shelf weakens though the next redshift bin, leaving nothing more than an extended tail of the red sequence
  by $0.5{\le}\,z{<}0.65$ (Fig.~\ref{d4000_dist}c). The first
  feature which could be fitted as a blue cloud population at
  $0.5{\le}\,z{<}0.65$ is the weak
  shelf at d$4000\,{\sim}1$.4--1.55 in the \smash{$10.8\,{\le}\log_{10}\!\mathcal{M}{<}11.0$}
  mass bin. In fact, the d4000 distribution of the
  11.0--11.2 stellar mass bin at $0.80{<}z{<}1.0$, most closely
  matches that in the 10.8--11.0 mass bin at $0.65{<}z{<}0.80$ and the
  10.6--10.8 mass bin (orange curve) at $0.50{<}z{<}0.65$. 
 The upper stellar mass limit of the blue cloud
  appears thus to drop by ${\sim}0.3$\,dex from
  \smash{$\mathcal{M}\,{\sim}10^{11.2\,}{\rm M}_{\odot}$} at $z{\sim}0.90$ to
  \smash{$\mathcal{M}\,{\sim}10^{10.9\,}{\rm M}_{\odot}$} by $z{\sim}0.58$. 
An examination of Fig.~\ref{d4000_dist}d indicates that this mass
limit drops by a further 0.2\,dex by the present day, the first
signature of the blue cloud being the shelf apparent at
d$4000\,{\sim}1$.5--1.7 in the 10.6--10.8 mass bin (orange curve). 

\begin{figure}
  \centering
  \includegraphics[clip, trim=0.1cm 2.2cm 2.0cm 3.3cm, width=8.4cm]{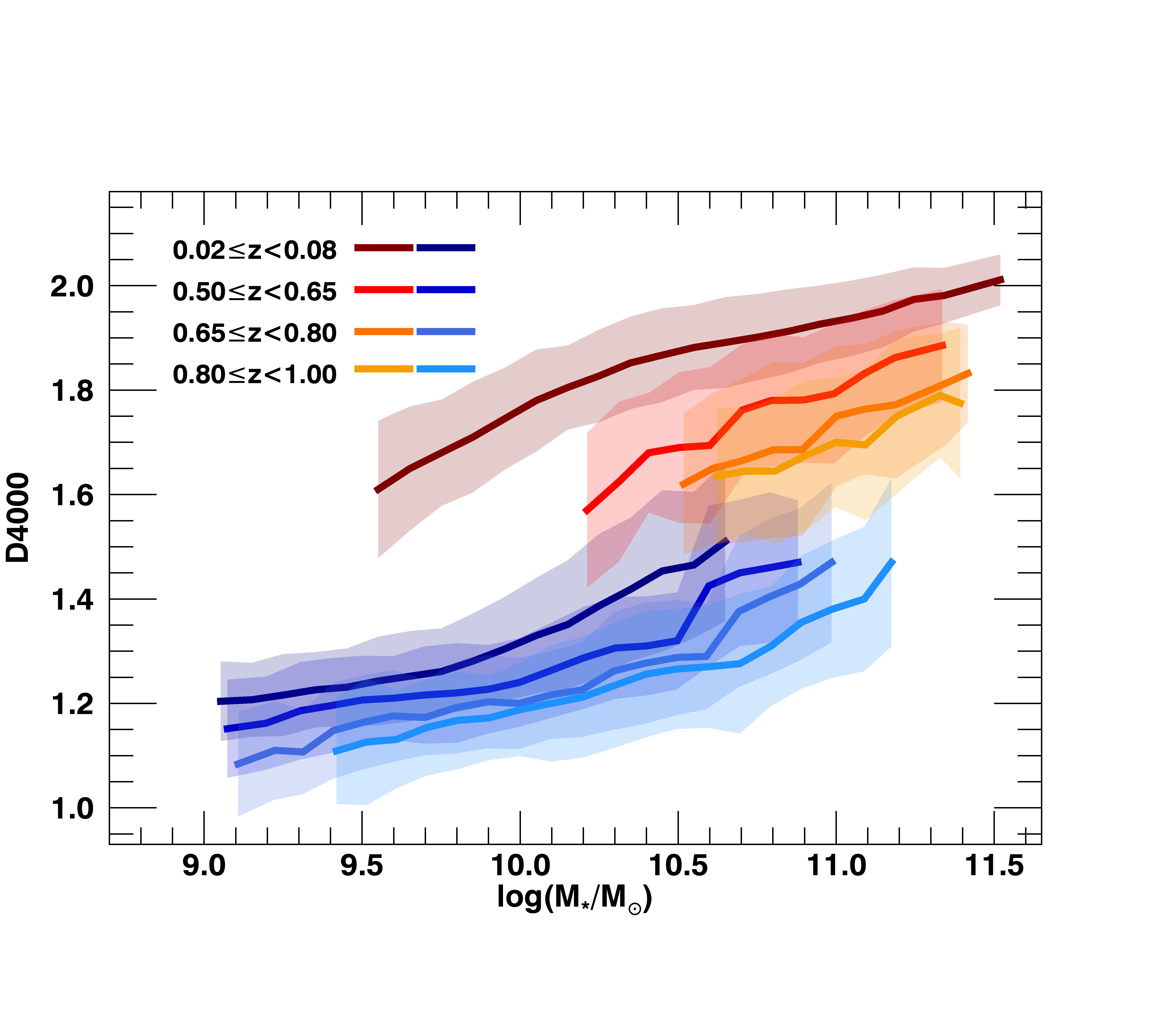}
\caption{Evolution of the blue cloud and red sequence in the \mbox{d4000--$\mathcal{M}$} diagram from $z{=}1.0$ to the present day. Each solid
curve indicates the central d4000 value of the Gaussian fit to either the
blue cloud or red sequence, as a function of stellar mass. The
shaded regions indicate the 1-$\sigma$ widths of each Gaussian fit. }
\label{d4000seq_evolution}
\end{figure} 

This evolution of the blue cloud is summarized in
Figure~\ref{d4000seq_evolution}, which plots the central d4000 values
and 1$\sigma$ widths 
of the blue cloud and the red sequence populations as a function of
stellar mass for the four redshift bins discussed above. 
This shows that the location and extent of the blue cloud has evolved
smoothly and continuously between $z{\sim}1$ and the present day in
two ways. First, the upper stellar mass limit of the
blue cloud has retreated steadily by ${\sim}0.5$\,dex
from \smash{$\log_{10}\!\mathcal{M}{\sim}11.2$} at $z{\sim}0.9$ down to 10.7 at $z{\sim}0.06$, as described
above. Second, the whole blue cloud has shifted to higher d4000
values at fixed stellar mass. For \smash{$\mathcal{M}\,{\la}10^{10.5\,}{\rm M}_{\odot}$}, 
where the blue cloud has a Gaussian-like d4000
distribution, the central d4000 values increase by 0.05--0.07 between the 0.8--1.0 and
0.50--0.65 redshift bins, and another 0.07--0.10 by the present day.
This gradual increase in d4000 with time probably reflects a combination
of the steady ${\sim}10{\times}$ decline in the specific SFRs of the star-forming main
sequence between $z{\sim}1$ and the present day \citep{noeske,behroozi,lee}, and a slow ageing of
their stellar populations over the intervening 8\,Gyr.

In each redshift bin, the d4000--$\mathcal{M}$ relation for the blue cloud appears approximately
linear with a slope of ${\sim}0$.12--0.14 (with no sign of
evolution), that starts to deviate from the linear approximation by
turning upwards towards the high-mass limit. 
This upturn appears to initiate at ever lower stellar masses as time
advances, from $\mathcal{M}{\sim}10^{10.7\,}{\rm M}_{\odot}$ at
$z{\sim}0.9$ to $\mathcal{M}{\sim}10^{10.2\,}{\rm M}_{\odot}$ by the present day. 

\subsection{Evolution of the red sequence since $z{\sim}1$}
\label{sec_redseq}

The evolution of the red sequence from $z{\sim}1$ to the present day
shown by Figure~\ref{d4000seq_evolution} can be summarized by two main changes. 
First, the red sequence as a whole has steadily shifted  
upwards to larger d4000 values with time, by $\sim$0.12
between the 0.8--1.0 and 0.50--0.65 redshift bins, and a further
${\sim}$0.13 by the present day. The d4000--\smash{$\mathcal{M}$} relation appears roughly
linear with a slope ${\sim}0.23$ which does not vary within the VIPERS
data, but flattens slightly to ${\sim}0.15$ for the local SDSS dataset.
The d4000--\smash{$\mathcal{M}$} relation for quiescent galaxies in VIPERS and its meaning in
terms of the evolution of their stellar populations is beyond the
scope of this paper, and is the focus of \citet{siudek}. 
 
Second, while the red sequence
already appears to be in place at the highest stellar masses at
$z{\sim}0.9$, it can be seen to 
extend to ever lower stellar masses with time, reaching
\smash{$\mathcal{M}\,{\sim}10^{9.6\,}{\rm M}_{\odot}$} by the present day. While the lower extents of the
red sequence in the VIPERS data are essentially determined by the
$i{=}22.5$ magnitude limit of the survey, corresponding to
$\mathcal{M}\,{\sim}10^{10.2\,}{\rm M}_{\odot}$ at $z{=}0.65$ for quiescent galaxies, the red sequence is only partly in
place in the 10.2--10.4 stellar mass bin immediately above this limit
at $0.5{\le}\,z{<}0.65$, appearing as no more than a shelf at
d400$0\,{\sim}1$.5--1.7 extending out of the dominant blue cloud (Fig.~\ref{d4000_dist}c). 

 Figure~\ref{d4000_dist} shows that the stellar mass bins which
first show a Gaussian-like d4000 distribution for the red sequence
population are those at the highest stellar masses. Only the \smash{$\mathcal{M}\,{>}10^{11.2\,}{\rm
  M}_{\odot}$} bin can be considered as being even marginally in place at
0.80--1.00 (Fig.~\ref{d4000_dist}a), while the first two mass bins (\smash{$\mathcal{M}\,{>}10^{11.0\,}{\rm
  M}_{\odot}$}) appear Gaussian-like at 0.65--0.80
  (Fig.~\ref{d4000_dist}b), increasing to the top three mass bins (\smash{$\mathcal{M}\,{>}\,10^{10.8\,}{\rm
  M}_{\odot}$}) at $0.5{\le}\,z{<}0.65$ (Fig.~\ref{d4000_dist}c). The
  finding of strongly-peaked d4000 distributions are indicative of
  relatively {\em mature} red sequence populations at these high stellar masses. 
   Newly quenched galaxies (with d400$0{\sim}1$.6--1.7) arrive onto
  the sequence at a rate that is much lower than that when the bulk of
  the sequence was put in place. Instead, the stellar mass ranges
  where the d4000 distribution appears rather flat over 1.4--1.8,
  should be those where the red sequence is {\em immature}, and being
  assembled at that redshift through the ongoing quenching of galaxies
  previously on the blue cloud.  This can be seen to be occurring at
  \smash{$10.6\,{\la}\log_{10}\!\mathcal{M}\,{<}11.2$} at $0.8{\le}\,z{<}1.0$
  (Fig.~\ref{d4000_dist}a), \smash{$10.4\,{\la}\log_{10}\!\mathcal{M}\,{<}11.0$} at
  $0.65{\le}\,z{<}0.80$ (Fig.~\ref{d4000_dist}b) and \smash{$10.2\,{\la}\log_{10}\!\mathcal{M}\,{<}10.8$} at
  $0.50{\le}\,z{<}0.65$ (Fig.~\ref{d4000_dist}c). In other words, the
  stellar mass range where the red sequence is in the process of being assembled shifts
  steadily downwards with time. 

\subsection{Evolution in the transition mass of galaxies}
\label{sec_transition}

One way of quantifying the effect of downsizing is to measure the
transition mass, \smash{$\mathcal{M}_{\rm cross}(z)$}, of galaxies as a function of
redshift. This is defined as the stellar mass at which the number densities of blue cloud and red
sequence galaxies are equal, and their stellar mass functions
intersect. It marks the transition from low-mass galaxies that are
predominately star-forming to high-mass systems that are mostly
quiescent \citep{k03}. 

\begin{figure}
  \centering
  \includegraphics[width=8.0cm]{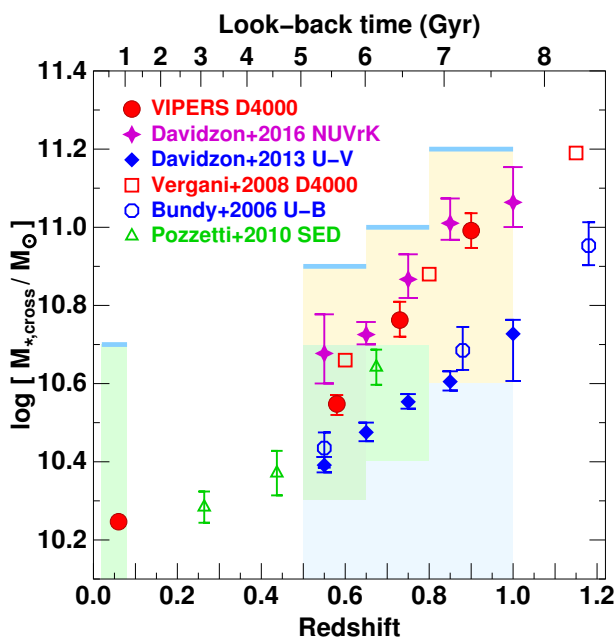}
\caption{Evolution of the transition mass $\mathcal{M}_{\rm cross}$ as a function
  of redshift. Red solid circles mark the stellar masses where the
  number densities of blue cloud (d$4000{<}1.55$) and red sequence
  (d$4000{>}1.55$) galaxies are equal, for each of the four redshift
  bins. Solid points are based on VIPERS data, while open points are
  $\mathcal{M}_{\rm cross}$ values from the literature. Blue diamonds are
  the $\mathcal{M}_{\rm cross}$ values from \citet{davidzon13} using $U-V$ colour to
classify blue and red galaxies, while magenta stars are the
$\mathcal{M}_{\rm cross}$ values obtained applying the NUV$rK$ classification of
\citet{davidzon16} to the same VIPERS-PDR1 sample. The points from
\citet{vergani08} are based on classification by D4000 break (red
squares), those from \citet{bundy} are based on a $U-B$
classification (blue octagons), and those of \citet{pozzetti} are based on a best-fit
SED classification (green triangles). Light blue horizontal lines mark
the upper mass limit of the blue cloud, pale yellow shaded regions
mark the stellar mass ranges where the d4000 distribution appears
flat-topped, the pale green shaded regions indicate the stellar mass
ranges where a ``Green Valley'' is apparent as a central dip in the d4000
distribution, while the light blue shaded regions are dominated by the
blue cloud population.}
\label{transition}
\end{figure}

 We estimate \smash{$\mathcal{M}_{\rm cross}$} for each of our four
redshift bins, where the number densities of blue cloud
(d$4000\,{<}1.55$) and red sequence (d$4000\,{>}1.55$) galaxies are equal,
as shown in Fig.~\ref{transition} by the solid red circles.
The transition mass falls steadily with time within the VIPERS dataset
from $\log_{10}\!\mathcal{M}_{\rm cross}{=}10.99$ at $z{=}0.90$ to
$\log_{10}\!\mathcal{M}_{\rm cross}{=}10.55$ 
at $z{=}0.58$. The latter is very close to the $3{\times}10^{10\,}{\rm
  M}_{\odot}$ value found by \citet{k03} for local galaxies,
suggesting little further evolution since $z{\sim}0.5$. However, this
oft-repeated value was only provided as a ballpark figure in \citet{k03}
where many global galaxy properties were seen to change. Applying the
exact same threshold of d$4000{=}1.55$, we find $\log_{10}\!\mathcal{M}_{\rm cross}{=}10.25$ or
$1.76{\times}10^{10\,}{\rm M}_{\odot}$ for SDSS galaxies with
$0.02{\le}z{<}0.08$. For comparison, \citet{moustakas} obtain $\log_{10}\!\mathcal{M}_{\rm
  cross}{=}10.33$ from their $0.01{\le}z{<}0.20$ SDSS-GALEX sample.

The $\mathcal{M}_{\rm cross}$ values from our VIPERS analysis are fully consistent with
those produced by \citet{vergani08} applying a similar d4000
classification (red squares in Fig.~\ref{transition}) to galaxies in the VIMOS VLT Deep Survey
\citep[VVDS;][]{lefevre}. Their results indicate that the transition mass keeps
increasing steadily beyond $z{\sim}1$, reaching $\mathcal{M}_{\rm
  cross}{=}10^{11.19\,}{\rm M}_{\odot}$ at $z{\sim}1.15$. 

The transition masses presented here are systematically
${\sim}0$.2--0.3\,dex higher than those of \citet{davidzon13} who applied a
$U{-}V$ classification to the VIPERS PDR1 sample (blue
diamonds). However, when classifying galaxies using the NUV$rK$
diagnostic of \citet{davidzon16}, and reperforming the analysis on the
same PDR1 sample, the resultant transition masses (magenta stars) and trends now
appear rather consistent with those obtained using our d4000 classification.
\citet{bundy} split galaxies from the DEEP2 redshift survey by their
($U{-}B$) colour, obtaining $\mathcal{M}_{\rm cross}(z)$ values (open blue octagons)
in agreement with those of \citet{davidzon13}. The lower transition masses obtained when using optical
colours to split galaxies could be due to the significant
numbers of star-forming galaxies known to contaminate the red sequence
due to the effects of dust reddening \citep{haines08,brammer11}. This is
supported by the $\mathcal{M}_{\rm cross}(z)$ obtained using the NUV$rK$
classifier being in much better agreement with those obtained using
the d4000 spectral feature. 

\begin{figure}
  \centering
  \includegraphics[height=8.0cm]{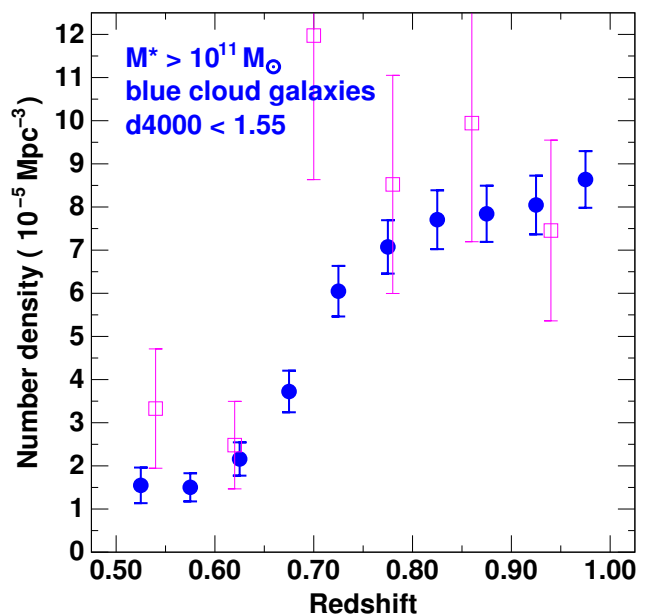}
\caption{Evolution in the number density of massive star-forming
  galaxies ($\mathcal{M}{>}10^{11\,}{\rm M}_{\odot}$ and d$4000{<}1.55$) 
from VIPERS (blue solid points) and the zCOSMOS
20K bright galaxy sample (magenta open squares) over $0.5{\le}z{<}1.0$.}
\label{M11blue_evolution}
\end{figure} 

To link these transition masses back to the d4000 distributions of
Fig.~\ref{d4000_dist}, the upper stellar mass limits of the blue cloud
determined in {\S}\,\ref{sec_bluecloud} are shown in
Fig.~\ref{transition} by blue horizontal lines.
The steady fall in \smash{$\mathcal{M}_{\rm cross}$} at least keeps pace (and may be more
rapid) with the retreat of the upper stellar mass limit of the blue
cloud since $z{\sim}1$.
For the upper reaches of the blue cloud at $0.5{\le}z{<}1.0$, the
d4000 distribution appears unimodal, with a flat top extending over
over d4000 values 1.4--1.8. The stellar mass bins where this
flat-topped distribution is seen are shown in Fig.~\ref{transition} by
the yellow shaded regions. Interestingly the lowest stellar mass where
this behaviour is seen shows no evidence of evolving, remaining at ${\sim}10^{10.7\,}{\rm
  M}_{\odot}$. This is approximately the characteristic stellar mass $\mathcal{M}_{SF}^{*}$
obtained for the stellar mass function of star-forming galaxies, and
which has been found to remain invariant at
$\log_{10}(\mathcal{M}_{SF}^{*}){\sim}10.65$ over $0.0\,{\le}\,z\,{\la}1.5$ in
numerous studies \citep[e.g.][]{peng10,ilbert13,tomczak,moutard}. This
supports the view that $\mathcal{M}_{SF}^{*}$ marks the critical scale above which
mass quenching becomes widepread \citep{bell07}. The stellar mass
range where a Green Valley can be discerned as a central dip between
relatively distinct red and blue sequences (green shaded region in
Fig.~\ref{transition}), can be seen to extend downwards from ${\sim}\mathcal{M}_{SF}^{*}$
to ever lower stellar masses with time, paralleling the evolution in
$\mathcal{M}_{\rm cross}$ and the high-mass limit of the blue
cloud.

\section{Evolution at the high-mass limit of the blue cloud}
\label{sec_m11}

The previous section showed how the high-mass limit of the blue cloud
has retreated from \smash{${\mathcal M}\,{\sim}10^{11.2\,}{\rm M}_{\odot}$} at
  $z{\sim}0.90$ to \smash{${\mathcal M}\,{\sim}10^{10.9\,}{\rm M}_{\odot}$} by $z{\sim}0.58$. 
How does this retreat (downsizing) translate into an evolution in the actual number
density of massive galaxies? 

Figure~\ref{M11blue_evolution} plots the evolution in the number
density of massive (\smash{$\mathcal{M}\,{\ge}10^{11\,}{\rm
  M}_{\odot}$}) blue-cloud galaxies with d$4000{<}1.55$ (blue points)
over $0.50{\le}z{<}1.0$. 
The volume covered by VIPERS is sufficient to contain over 1,000
massive blue-cloud galaxies over this redshift range, allowing it to
be finely divided into narrow redshift bins just 0.05 in width. Each
galaxy is weighted according to its TSR and SSR to account for
spectroscopic incompleteness. Table~1 reports the
numbers of VIPERS galaxies in each redshift bin, and the resultant
number density of massive blue-cloud galaxies.
Based on their observed apparent $i$-band
magnitudes, the loss of massive blue-cloud galaxies from the VIPERS
survey due to being too faint ($i_{AB}{>}22.5$) should be negligible, at
least up to $z{\sim}0.9$. We also do not expect to miss massive blue
cloud galaxies due to the VIPERS $u-g$, $g-i$ colour selection
criteria, as their observed colours remain reasonably far from the
colour cuts used. 

\begin{figure}
  \centering
  \includegraphics[height=8.0cm]{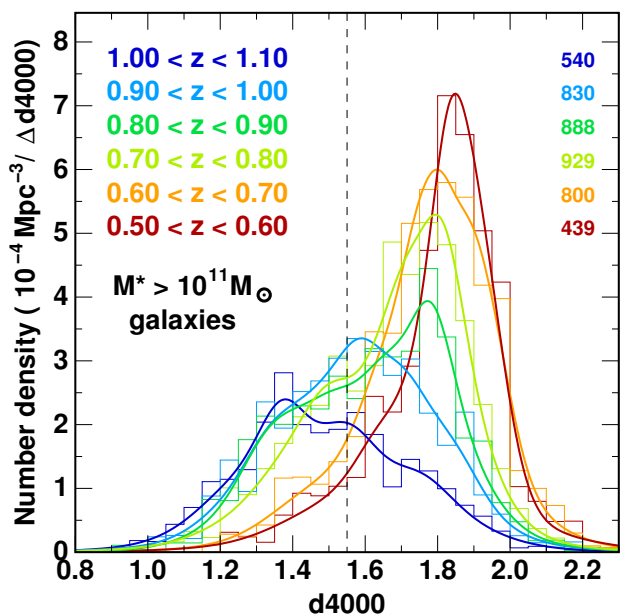}
  \caption{Evolution of the d4000 distribution of massive
    galaxies. The curves are normalized to represent the number of
    $\mathcal{M}\,{>}10^{11}\,{\rm M}_{\odot}$ galaxies per unit
    comoving volume per unit d4000.}
\label{M11_d4000evol}
\end{figure} 
 
The number density of
massive blue-cloud galaxies remains relatively constant at a value
${\sim}8{\times}10^{-5\,}{\rm Mpc}^{-3}$ over
$0.75{\le}z{<}1.0$, before rapidly dropping by a factor 4--5 over a
time-scale of just ${\sim}1$\,Gyr, to reach a value of just
${\sim}1.5{\times}10^{-5\,}{\rm Mpc}^{-3}$ by $z{\sim}0.6$. 
We find 342 massive blue-cloud galaxies within our SDSS-DR7 subsample
over $0.02{<}z{<}0.08$, corresponding to a number density of
 $1.34{\pm}0.15{\times}10^{-5\,}{\rm Mpc}^{-3}$. This suggests no
 further significant decline in the population after $z{\sim}0.5$. 

The sharp decline over $0.5{<}z{<}0.8$ is also seen in the zCOSMOS data (magenta 
squares), taken from the final zCOSMOS-bright sample of about 20\,000
galaxies over the 1.7\,deg$^{2}$ COSMOS field and selected simply
to have $I_{AB}{\le}22.5$ \citep{lilly07,lilly09}, as measured from the COSMOS
{\em HST}-ACS F814W images. Each galaxy is weighted to account for
local spectroscopic incompleteness in the zCOSMOS survey in the same
way as VIPERS. 
The smaller areal coverage means that cosmic variance effects are
much more important for zCOSMOS, and the error bars include the expected
bin-to-bin uncertainties due to cosmic variance (20--25\% for our redshift
bins of width 0.08), assuming the formula of \citet{driver}. 
The number densities seen in the three highest redshift bins in the
zCOSMOS data are consistent with the high plateau seen in VIPERS,
while the two lowest redshift bins show number densities
${\sim}3{\times}$ lower, close to those seen in the much larger volume
covered by VIPERS. The only inconsistency is seen in the bin centred at $z{=}0.7$, which
contains two previously known wall-like structures of connected groups
at $z{\sim}0.67$ and $z{\sim}0.73$ \citep{knobel,iovino} that dominate the number counts for this bin. 
The largest source of uncertainty in the evolution seen in the VIPERS
data is unlikely to be Poisson noise or cosmic variance, but systematics in the
stellar mass estimates. \citet{moustakas} show that changing the 
stellar population synthesis models or the priors (e.g. the
stochastic use of bursts) can affect the level of evolution observed in
the number densities of massive star-forming galaxies (their Fig.~20). 
However, these effects do not appear large enough (${\la}0$.2\,dex
over $0{<}z{<}1$) to fundamentally change our result. 

The natural question at this point is where have these massive blue
cloud galaxies gone? Assuming that they haven't lost any stellar mass,
they must have left the massive blue cloud sample by becoming
quiescent, increasing their d4000 values beyond our limit of 1.55. 
Figure~\ref{M11_d4000evol} plots the d4000 distributions of VIPERS
galaxies with $\mathcal{M}\,{>}10^{11\,}{\rm M}_{\odot}$ normalized
per unit comoving volume, in six
redshift bins spanning $0.5{\le}z{<}1.1$ 

\begin{figure}
  \centering
  \includegraphics[height=8.0cm]{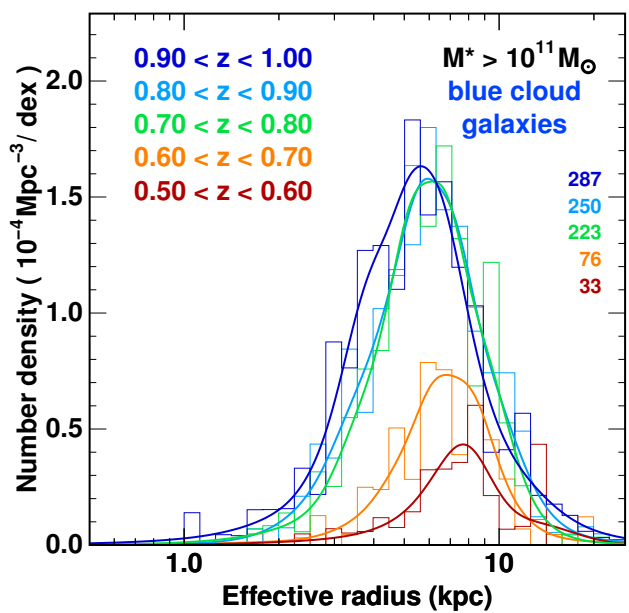}
\caption{Evolution in the size distribution of massive star-forming
  galaxies with $\mathcal{M}\,{>}10^{11\,}{\rm M}_{\odot}$ and
  d$4000{<}1.55$ over $0.5{\le}z{<}1.0$, normalized per unit comoving
  volume per dex in $r_{e}$.}
\label{size_evolution}
\end{figure} 

 This figure encapsulates the large-scale shift in the d4000
distribution of the most massive galaxies from being a dominant blue
cloud population peaking at d$4000{\sim}1.35$ at $1.0{\le}z{<}1.1$,
the peak shifting through intermediate values (d$4000{\sim}1.6$) in
the next redshift slice (0.9--1.0), before the red sequence becomes
gradually more dominant through to $z{=}0.5$. It is clear that the
majority of the massive galaxies which resided on in the blue cloud at
$z{\sim}1.1$ have been transferring steadily 
through the green valley and onto the red sequence by
$z{\sim}0.5$. 
These results are fully consistent with the systematic shifts in the
d4000 distributions of $\mathcal{M}\,{>}10^{11\,}{\rm M}_{\odot}$ galaxies
from the VIMOS VLT Deep Survey (VVDS), with the median d4000 value shifting from 1.37 at
$1.0{\le}z{<}1.3$ to 1.69 at $0.5{\le}z{<}0.7$ \citep[Fig.~1 of][]{vergani08}.

\begin{figure*}
  \centering
  \includegraphics[width=18.4cm]{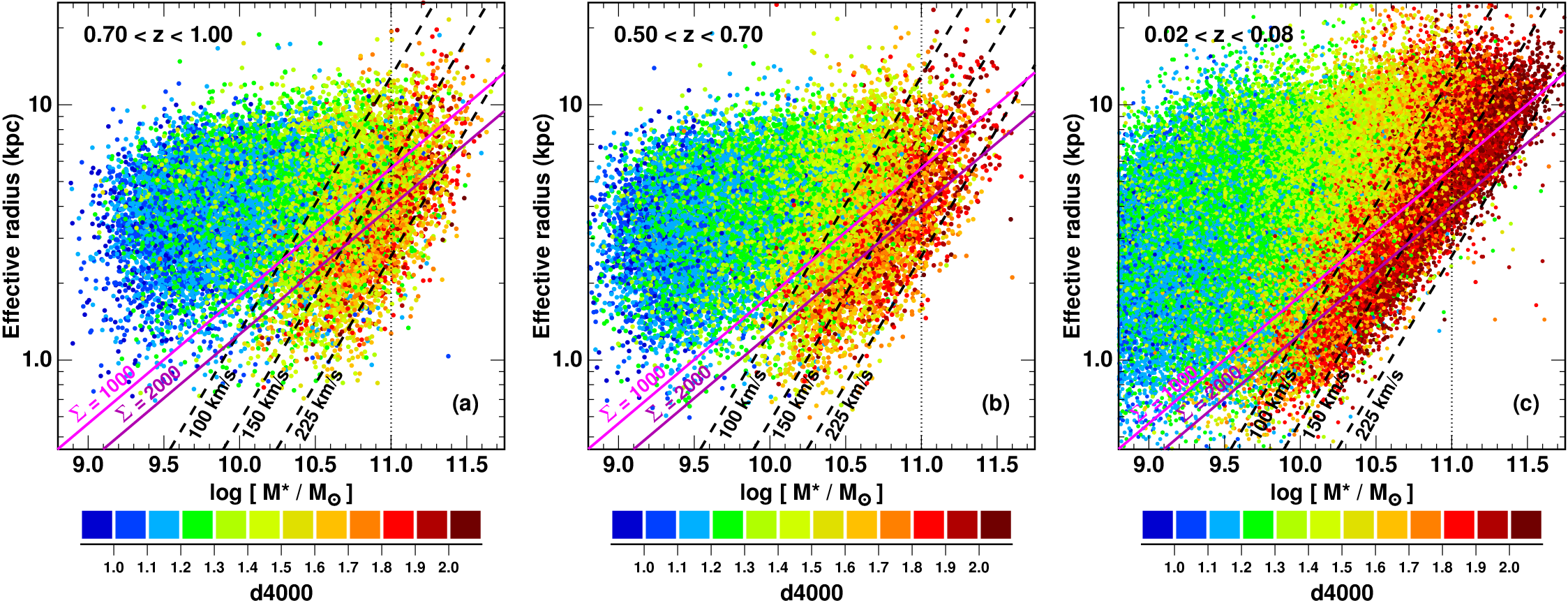}
  \caption{Evolution of the size-mass relation of galaxies, plotting
    effective radius versus stellar mass for three
    redshift ranges: 0.70--1.00, 0.50--0.70 (VIPERS) and 0.02--0.08
    (SDSS data). Each galaxy is colour coded according to its d4000 value
    as indicated in the colour bars along the bottom. Black diagonal
    dashed lines mark lines of constant inferred velocity dispersion
    of 100, 150 and 225\,km\,s$^{-1}$ following \citet{franx}. Magenta
    lines indicate effective stellar mass surface densities $\Sigma_{e}{=}1$000 and 2000\,M$_{\odot}$\,pc$^{-2}$ as used in \citet{gargiulo}.}
\label{size_mass_evolution}
\end{figure*}

\begin{table}
\caption{Number density evolution of massive blue-cloud galaxies}
\centering
\begin{tabular}{ccr@{$\,\pm\,$}l} \hline
Redshift & $N_{\rm gals}$ & \multicolumn{2}{c}{$\rho_{\rm gals}$} \\
range & $\mathcal{M}\,{>}10^{11\,}{\rm M}_{\odot}$ &
\multicolumn{2}{c}{($10^{-5\,}{\rm Mpc}^{-3}$)} \\ \hline
0.50--0.55 & 17   & 1.55 & 0.42 \\
0.55--0.60 & 24   & 1.50 & 0.33 \\
0.60--0.65 & 34   & 2.16 & 0.39 \\
0.65--0.70 & 67   & 3.72 & 0.49 \\
0.70--0.75 & 121 & 6.05 & 0.59 \\
0.75--0.80 & 146 & 7.08 & 0.63 \\
0.80--0.85 & 161 & 7.71 & 0.67 \\
0.85--0.90 & 176 & 7.84 & 0.67 \\
0.90--0.95 & 175 & 8.05 & 0.70 \\
0.95--1.00 & 202 & 8.64 & 0.65 \\
1.00--1.05 & 177 & 8.08 & 0.71 \\ \hline
\end{tabular}
\end{table}

Figure~\ref{size_evolution} presents the size distributions of these massive
blue cloud ($\mathcal{M}\,{>}10^{11\,}{\rm M}_{\odot}$, d$4000{<}1.55$)
galaxies, for five redshift bins covering $0.5{\le}z{<}1.0$, where
size is quantified by the half-light radius, $r_{e}$, in kpc. 
The $y$ axis indicates the number density of galaxies per unit
comoving volume per dex in $r_{e}$, so that vertical shifts between
curves reflect changes in the number density of blue cloud galaxies of
a fixed size. 
The highest three redshift bins are all consistent with a log-normal
distribution of mean $r_{e}{=}5.85$\,kpc and 0.18\,dex width, with no
apparent changes in the number density at fixed size. 
The mean proper effective radius increases with time for the lowest two
redshift bins, as the overall number density sharply drops, to $\langle
r_{e}\rangle{=}7.39$\,kpc for $0.5{\le}z{<}0.6$. The likelihood that a
massive galaxy that was on the blue cloud at $z{\sim}0.8$ and is still
there at $z{\sim}0.5$, can be seen to depend strongly on its
size. While the number density of massive blue cloud galaxies more
compact than average ($r_{e}{<}5.85$\,kpc) drops by a factor ten
between $z{\sim}0.8$ and $z{\sim}0.5$, it falls by just a factor three
for those galaxies larger than average. At the same time, the mean
size of those massive galaxies which must have left the blue cloud
between $z{\sim}0.8$ and $z{\sim}0.5$ is still 5.5\,kpc.

In contrast we find little evidence for evolution in the
S\'{e}rsic-index distribution of massive blue-cloud galaxies over this period, or a preferential
quenching of these galaxies based on their structure. The distribution extends over
$0.8{\le}\eta{\le}5.0$ (i.e. spanning both classical exponential and de
Vaucouleurs profiles), peaking at intermediate
values ($\eta{\sim}2$) at all redshifts covered by VIPERS (0.5--1.0).

\section{Evolution of the size-mass relation}
\label{sec_sizemass}

 We examine the evolution of the size-mass relation of galaxies from
$z{\sim}1$ to the present day in Figure~\ref{size_mass_evolution},
plotting effective radius versus stellar mass for three
redshift intervals: 0.7--1.0, 0.5--0.7 (VIPERS) and 0.02--0.08
(SDSS). 
The VIPERS data are now split into just two
redshift bins $0.7{\le}z{<}1.0$ and $0.5{\le}z{<}0.7$, which
encapsulate the two phases of evolution seen in
Fig.~\ref{M11blue_evolution}, and span
similar look-back times (1.41 and 1.26\,Gyr respectively).
Galaxies are colour-coded by their d4000 values, enabling the 
dependence of a galaxy's star-formation history on its size and
stellar mass to be separated objectively, without {\em a priori}
imposing a specific combination of the two properties
(e.g. \smash{$\mathcal{M}/r_{e}$}, $\Sigma_{e}$). Black diagonal dashed lines mark constant inferred velocity
dispersions of 100, 150 and 225\,km\,s$^{-1}$ assuming the relation
$\sigma_{\rm inf}{=}\sqrt(0.3 G\mathcal{M}/r_{e})$
 of \citet{franx}, where the constant was chosen so that the
 $\sigma_{\rm inf}$ of the SDSS galaxies matched the measured velocity
 dispersions \citep[see also][]{bezanson}. 

\begin{figure*}
  \centering
  \includegraphics[width=18.4cm]{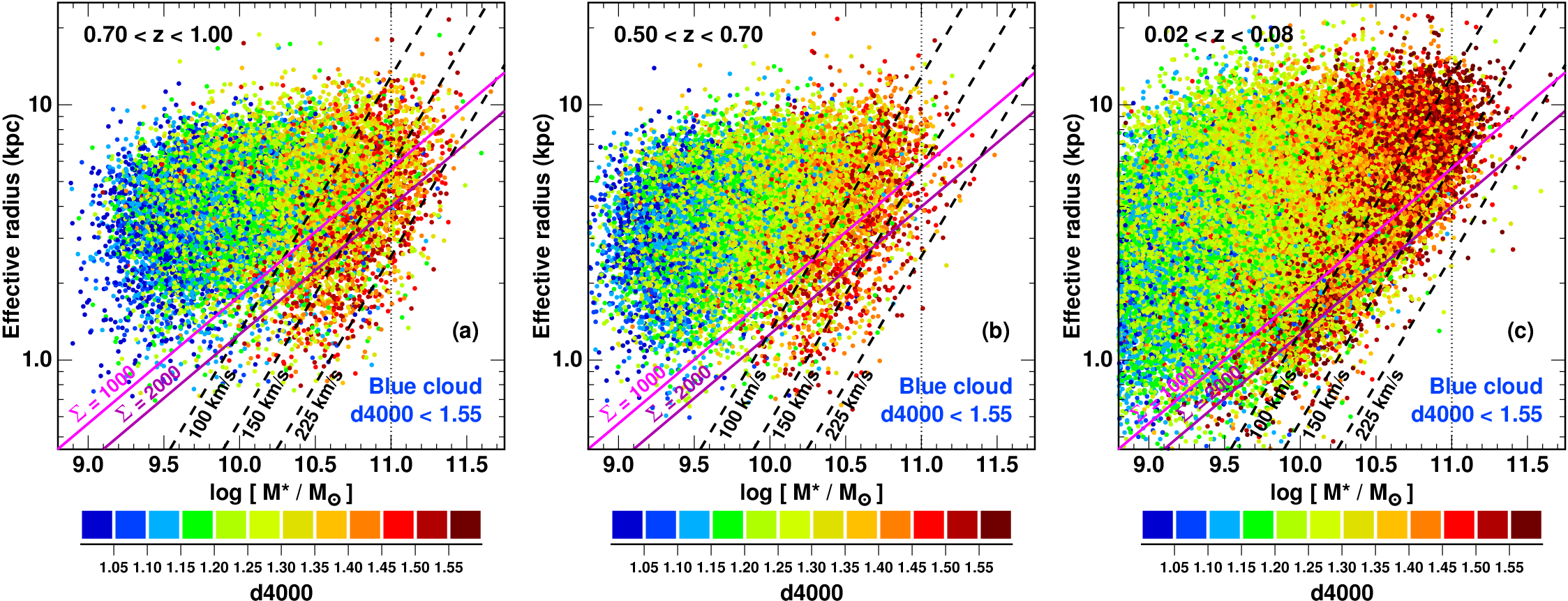}
  \caption{The size-mass relation of blue cloud galaxies
    (d4000$<$1.55), plotting effective radius $r_{e}$ versus $\mathcal{M}$ for
  the same three redshift ranges as in Figure~\ref{size_mass_evolution}: 0.7--1.0, 0.50--0.70 (VIPERS) and
  0.02--0.08 (SDSS). Each galaxy is
    colour coded according to its d4000 value, as indicated in the
    colour bar along the bottom. Black dashed lines mark 
    inferred velocity dispersions of 100, 150 and
    225\,km\,s$^{-1}$.}
\label{size_mass}
\end{figure*}

 Figure~\ref{size_mass_evolution}c shows $0.02{<}z{<}0.08$ galaxies from the SDSS-DR7 sample.
Only galaxies identified as centrals by \citet{yang} are
plotted. Hence any trends should be due to the effects of mass
quenching rather than environmental quenching due to the galaxy
becoming a satellite within a more massive halo.
Most notable is that the galaxy population is very sharply bounded on
three sides of the size-mass plot. There is an upper size limit of
10\,kpc, rising slowly with stellar mass to 20\,kpc by
$\mathcal{M}\,{\sim}10^{11\,}{\rm M}_{\odot}$, and a lower boundary at
${\sim}0.7$\,kpc (which may be a resolution limit). Most importantly,
there is a very sharp upper stellar mass limit, which runs diagonally
through the size-mass plot, approximately along the line of constant
velocity dispersion at $\sigma_{\rm inf}{=}225$\,km\,s$^{-1}$. The empty
region to the lower-right of this line is termed the Zone-of-Exclusion
by \citet{cappellari}, as the region of the size-mass relation where
none of their 260 local early-type galaxies from the ATLAS$^{\rm 3D}$ survey
are found. \citet{taylor10} warn that the absence of
compact high-mass systems is at least partly due to SDSS spectroscopic
target selection criteria, which exclude objects on the basis of
having bright fiber magnitudes ($m_{\rm fib}{<}15$) to avoid cross-talk
between the fibers. The SDSS should however be ${>}80$\% complete for
compact, massive galaxies at $z{\ga}0.06$.
It is also immediately apparent that the star-formation history of
galaxies is also fundamentally defined by this same diagonal limit,
with all the old, quiescent galaxies (d$4000\,{\ga}1.8$; red points)
bounded within a narrow diagonal strip, some 0.5\,dex wide in \smash{$\mathcal{M}$}, that runs parallel to this fundamental
 high-mass limit for galaxies. The quiescent galaxy population is
 essentially bounded within the two diagonal lines marking inferred
 velocity dispersions of 100\,km\,s$^{-1}$ and 225\,km\,s$^{-1}$,
 while virtually all galaxies to the left of these lines have
 young stellar populations with d$4000\,{\la}1.4$ (green, blue points).

The VIPERS data allow the evolution of these sharp fundamental boundaries and
behaviours to be followed back in time to $z{\sim}1$ in comparable detail. 
The most notable aspect of Fig.~\ref{size_mass_evolution} is the lack of difference
in the basic distributions of galaxies in the size-mass plane at the
three epochs. There are well-defined upper and lower size limits of
${\sim}1$\,kpc and 10\,kpc in all three panels. 
The upper mass bound continues to run parallel to the lines of
constant velocity dispersion, and the oldest galaxies with the highest
d4000 values (red points) remain limited to a narrow diagonal band
along this edge. 
While in the SDSS data, the most massive galaxies
appear essentially confined just within the 225\,km\,s$^{-1}$ line, in the VIPERS
data, some galaxies do appear to extend just beyond it
(by up to 0.1--0.2\,dex), suggesting a marginal shift since $z{\sim}1$. 
The fundamental result of Figure~\ref{size_mass_evolution} appears
that the basic properties of the size-mass relation were already in
place at $z{\sim}1$, and have only evolved gradually to the present day.

While the location of these massive quiescent 
galaxies in the size-mass plane appears
to evolve little, moving to higher redshifts their d4000 values do
appear to decline, indicative of younger
stellar populations. 
The evolution of these massive, passive galaxies is the
focus of \citet{gargiulo}, while in this work we explore
what causes a star-forming galaxy at the upper end of the blue
cloud to leave it and become quiescent. 
However in Fig.~\ref{size_mass_evolution}, the upper-limits of the
blue cloud and galaxies in the first stages of transformation are 
hidden, submerged under the dominant population of old, already
quiescent galaxies. 

To gain insights into what processes push a galaxy
from the upper end of the blue cloud, and fundamentally limit the
ability of galaxies to continue to grow through star formation, 
Figure~\ref{size_mass} displays the size-mass relation for the same
three redshift intervals as Fig.~\ref{size_mass_evolution},  
but plotting only blue cloud galaxies (d$4000{<}1.55$).
Galaxies are colour-coded according to
their d4000 as before, but the scale has been changed so that orange/red
points mark galaxies on the high-d4000 wing of the blue cloud
($1.4{<}{\rm d}4000{<}1.55$), and
hence most likely to be those about to leave it. 
The d4000 values of blue cloud galaxies increase steadily with
stellar mass, from ${\sim}1.1$ (blue points) at the lowest masses to
${\sim}1.5$ (red points) at the high-mass end. For stellar masses
above $10^{10}{\rm M}_{\odot}$ the dependence of d4000 appears tilted,
such that smaller galaxies have slightly higher d4000 values at fixed stellar
mass than larger ones. Moreover, the high-mass upper limit of the blue
cloud appears tilted, such that it appears to run parallel to the
diagonal black dashed lines marking constant inferred velocity
dispersion $\sigma_{\rm inf}$. 
For comparison, lines of constant stellar mass
($\mathcal{M}{=}10^{11}\,{\rm M}_{\odot}$, dotted line) and stellar mass density
$\Sigma_{e}$ (magenta lines) run at
significant angles with respect to the upper-limit of the blue cloud. 

There is evidence of evolution between the two highest redshift bins. 
At $0.7{\le}\,z{<}1.0$ (Fig.~\ref{size_mass}a) the upper limit of the blue cloud is approximately
demarcated by the final black dashed line at $\sigma_{\rm inf}{=}2$25\,km\,s$^{-1}$,  
including a significant population with $\mathcal{M}\,{>}10^{11\,}{\rm M}_{\odot}$.
By $0.5{\le}\,z{<}0.7$ (Fig.~\ref{size_mass}b), the upper limit of the blue
cloud has now retreated to the middle black dashed line
marking $\sigma_{\rm inf}{=}150$\,km\,s$^{-1}$, with
just a smattering of points beyond it. 
This upper bound does not show signs of further movement to the
present day (Fig.~\ref{size_mass}c), with blue cloud galaxies from the SDSS-DR7 extending up
to the $\sigma_{\rm inf}{=}1$50\,km\,s$^{-1}$ dashed line, beyond
which the number density drops abruptly.

\begin{figure}
  \centering
  \includegraphics[width=8.0cm]{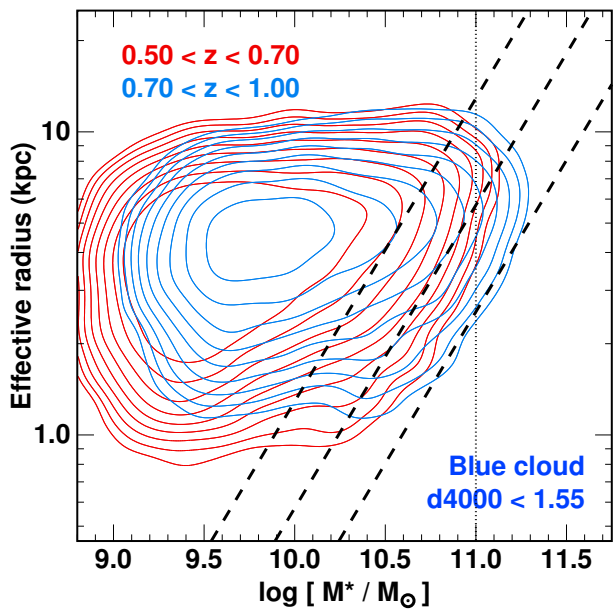}
\caption{Comparison of the bivariate number density distributions (per unit
  comoving volume) of effective radius ($r_{e}$) and stellar mass ($\mathcal{M}$)
for blue cloud galaxies (d$4000{<}1.55$) in the two redshift ranges: 0.7--1.0 (blue contours) and
0.5--0.7 (red contours). Each contour indicates a factor $\sqrt{2}$
change in density.}
 \label{contours}
\end{figure}

Figure~\ref{contours} directly compares the
bivariate number density distribution of galaxies in the two highest redshift
bins, weighting each galaxy to account for spectroscopic and stellar
mass incompleteness, and accounting for the different comoving volumes sampled
by VIPERS over the two redshift ranges. 
The high-mass limit of the blue cloud at $0.7{\le}\,z{<}1.0$ (blue contours)
can be seen to be shifted 0.17\,dex to higher stellar masses with
respect to that at $0.5{\le}\,z{<}0.7$ (red contours) at fixed effective
radius. The iso-density contours along this high-mass edge are 
parallel to the dashed lines for both redshift bins, confirming that
the upper limit of the blue cloud runs along lines of constant
$\mathcal{M}/r_{e}$.

At $0.7{<}z{<}1.0$ (Fig.~\ref{size_mass}a), the galaxies on the high-d4000 wing of the blue
cloud ($1.4{<}{\rm d}4000{<}1.55$; orange/red points) appear mostly confined to the
diagonal band between the first and third black dashed lines
($100{\le}\sigma_{\rm inf}{<}225$\,km\,s$^{-1}$). These are the galaxies
that are most likely to be in the process of leaving the blue cloud and become
the next arrivals onto the red sequence. 
A comparison to Fig.~\ref{size_mass_evolution}c shows that this band roughly corresponds to
the region enclosing massive passive galaxies at the present day.
Figure~\ref{size_mass}b indicates that by $0.5{<}z{<}0.7$, most of those galaxies with
$\sigma_{\rm inf}{>}150$\,km\,s$^{-1}$ must indeed have left the blue cloud,
while those with $100{<}\sigma_{\rm inf}{\la}150$\,km\,s$^{-1}$ now
represent the high-mass limit of the blue cloud, and having d4000
values ${\ga}1.40$, are those most likely to leave in the near future. 
This changes little to the present day (Fig.~\ref{size_mass}c). 

The extension of $1.4{<}{\rm d}4000{<}1.55$ galaxies to the left of the
$\sigma_{\rm inf}{=}100$\,km\,s$^{-1}$ lines at large radii
($r_{e}{\sim}10$\,kpc) may well be due to aperture effects, as the
3$^{\prime\prime}$ diameter (3.5\,kpc at $z{=}0.06$) SDSS fibres sample only inside
$r_{e}/4$ for such large galaxies. The presence of a bulge component and steep negative metallicity
gradients within disc galaxies could contribute to produce significant
biases in the d4000 measurements of these objects \citep{kewley,gonzalez}. 
At the redshift range of the VIPERS survey, the 1$^{\prime\prime}$
slit-width corresponds to 6.1\,kpc at $z{=}0.5$ and 8.0\,kpc at
$z{=}1.0$, and the slits are designed to be long enough to extend beyond the galaxy
in both directions in order to measure the sky background (unlike fibres).
We thus do not expect to be significantly affected by metallicity
gradients, and note that they should act in the opposite direction to
the observed trend. 

\begin{figure*}
  \centering
  \includegraphics[width=18.4cm]{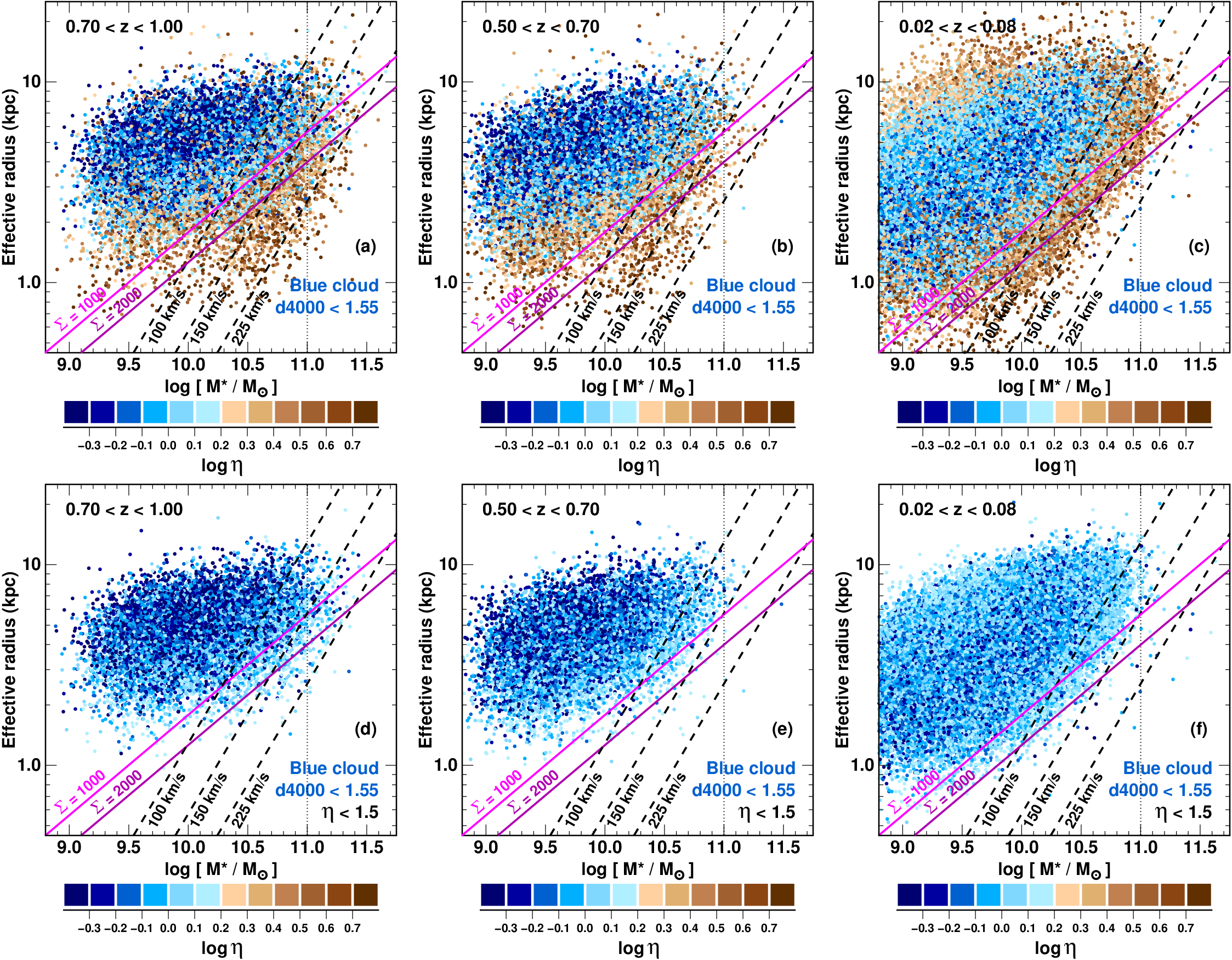}
  \caption{The size-mass relation of blue cloud galaxies (d4000$<$1.55), plotting effective radius $r_{e}$ versus $\mathcal{M}$ for
  three redshift ranges (upper panels): 0.7--1.0, 0.5--0.7 (VIPERS) and 0.02--0.08 (SDSS).
  Galaxies are colour coded according to their Sersic index, as indicated in the colour bar
  along the bottom. Blue points indicate late-type galaxies
  ($\eta{<}1.5$) while brown points are early-types. 
The three lower panels show the same size-mass relations
  but plotting only classic late-type galaxies with $\eta{<}1.5$.}
\label{size_mass_nser}
\end{figure*}

The band of blue cloud galaxies with the highest d4000 values
(1.40--1.55), while tilted, extends over the full range of effective
radii seen among the whole blue cloud population. The retreat in
the blue cloud revealed in Figure~\ref{contours} appears to be
systematic, with no significant dependence on effective radius. This suggests that
the size distribution of galaxies leaving the blue cloud is similar to
that of galaxies still in it.

\subsection{Changes in galaxy structure within the blue cloud}

\begin{figure*}
  \centering
  \includegraphics[width=18.4cm]{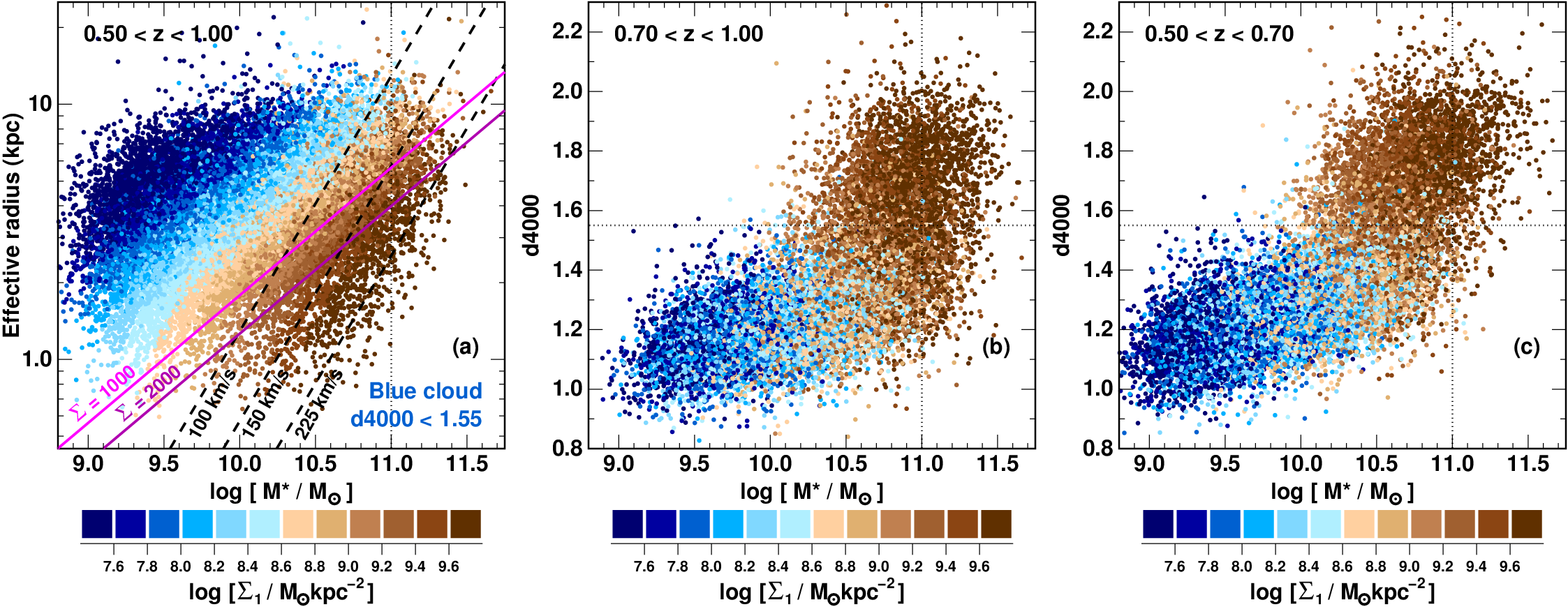}
  \caption{The impact of central stellar mass surface density
    $\Sigma_{1}$. The left panel plots the size-mass relation of
    blue cloud galaxies over $0.5{\le}z{<}1.0$, colour coded according
    to central stellar mass surface density $\Sigma_{1}$, as
    indicated in the colour bar along the bottom. The
    remaining panels plot the d4000 versus stellar mass relation for
    VIPERS galaxies for the redshift bins, 0.7--1.0 and 0.5--0.7,
    colour coded by $\Sigma_{1}$.}
  \label{smu_1kpc}
\end{figure*}

 The upper panels of Fig.~\ref{size_mass_nser} plot the size-mass relation of blue cloud
galaxies (d$4000{<}1.55$) for three different redshift intervals as
before, but now colour code galaxies according to their S\'{e}rsic index $\eta$ rather
than d4000 value. The first two panels show that the concentrations ($\eta$) of blue cloud
galaxies within VIPERS ($0.5{\le}z{<}1.0$) vary systematically with
size and stellar mass in two different directions. 
First, for galaxies to the left of
the $\sigma_{\rm inf}{=}100$\,km\,s$^{-1}$ line, their S\'{e}rsic index
increases steadily as galaxies become smaller, from
$\eta\,{\la}\,1.0$ for larger galaxies with $r_{e\,}{\ga}\,3$\,kpc, to
$\eta\,{\ga}\,2.5$ for the most compact systems with $r_{e\,}{\la}\,2$\,kpc, 
and little additional dependence on stellar mass. 
Second, the S\'{e}rsic index also increases with stellar
mass, with the $\sigma_{\rm inf}{=}100$\,km\,s$^{-1}$ line dividing
low-mass galaxies with late-type morphologies ($\eta{<}1.5$, blue
points) and high-mass systems with increasingly concentrated profiles,  
reaching $\eta\,{\ga}\,2.5$ (brown points) for those galaxies on the
high-mass edge of the blue cloud.
This latter trend parallels that seen for d4000 in
Fig.~\ref{size_mass}a,b for the same blue cloud galaxies. 
As galaxies in the blue cloud start to approach its
high-$\sigma_{\rm inf}$ edge, both their internal structures and their
ability to continue forming stars are being systematically affected,
even as they remain within the blue cloud.  
In contrast, no such change in d4000 values is seen for the compact low-mass
systems to parallel the increase in $\eta$. 
  
 The lower panels show that these same trends are apparent, even
when considering only secure late-type galaxies with
$\eta{<}1.5$. That is, the late-type galaxies with the highest
$\sigma_{\rm inf}$  (${>}100$\,km\,s$^{-1}$), also have the highest S\'{e}rsic indices ($1.0\,{\la}\,\eta{<}1.5$,
light blue points), while those to the left of the 100\,km\,s$^{-1}$ line have
$\eta\,{\la}\,1.0$ (darker blue points). Even the first phases of
structural transformation are aligned with an increase in
$\sigma_{\rm inf}$. 
The regions where late-type galaxies on the blue cloud can appear on
the size-mass plane can be seen to have shrunk considerably in
Figs.~\ref{size_mass_nser}d,e from the upper panels where no morphological
selection was made. First, all of the most compact galaxies
(${\la}1.5$\,kpc) have disappeared, reducing the overal radial (vertical) extent
of the blue cloud at all stellar masses.
More importantly, the high-mass limit appears to have retreated by
0.10--0.15\,dex. Only 30\% of $\sigma_{\rm inf}{>}100$\,km\,s$^{-1}$ blue cloud galaxies at
$0.5{<}z{<}0.7$ have $\eta{<}1.5$.  
This shrinkage demonstrates the comprehensive nature of the physical
processes that are driving the structural changes in blue cloud
galaxies in certain regions of the size-mass plane.

As before, the SDSS-DR7 dataset can be used to see whether the patterns
apparent in the size-mass relation of blue cloud galaxies at 
$0.5{\le}z{<}1.0$ remain to the present day. 
Figure~\ref{size_mass_nser}c confirms that  $0.02{\le}z{<}0.08$ blue cloud
galaxies are confined to the same regions of the
size-mass plane as VIPERS galaxies at $0.5{\le}z{<}0.7$. 
The high-mass limit of the blue
cloud at $z{\sim}0$ can be seen to essentially run along 
the line marking $\sigma_{\rm inf}{=}150$\,km\,s$^{-1}$. The
connections between morphology ($\eta$) and location of a galaxy in the
size-mass plane are also persistent. Low-mass galaxies with
$\sigma_{\rm inf}{<}100$\,km\,s$^{-1}$ are virtually all late types ($\eta{<}1.5$; blue points), except for the most compact 
($r_{e}{\la}1.5$\,kpc) and large systems ($r_{e}{\ga}10$\,kpc) which show higher concentrations. 
As blue cloud galaxies push through the $\sigma_{\rm inf}{=}100$\,km\,s$^{-1}$ line, and
approach the high-mass boundary of the blue cloud population, their
structures become increasingly concentrated ($\eta{>}2$; brown points).
Repeating this analysis using the bulge-to-total ratios of
\citet{simard} produces the same global trends
(Appendix~\ref{BTappendix}). Galaxies on the leading edge of the blue cloud have significant bulge components
($0.3{\la}{\rm (B/T)}{\le}1.0$), while lower mass systems are uniformly
disc-dominated (B/T${<}$0.25).

\subsection{The link to central stellar mass surface density}

Numerous studies have indicated that the central stellar mass surface density
$\Sigma_{1}$ is a much better predictor of star formation activity
within a galaxy than stellar mass \citep{cheung,fang,barro,whitaker16}, while the presence of a prominant
bulge appears a necessary condition for galaxies to quench star
formation on galaxy-wide scales \citep{bell08,bell12}. 
However, while the results
 presented above showing the clear trends connecting the changes in d4000
and $\eta$ with the location of galaxies in the size-mass relation,
and in particular along lines of constant \smash{$\mathcal{M}/r_{e}$} or
$\sigma_{\rm inf}$, it is not clear how these trends relate to $\Sigma_{1}$. 

Figure~\ref{smu_1kpc}a shows the variation of central
stellar mass surface density $\Sigma_{1}$ within the size-mass
plot for blue cloud galaxies, over the combined VIPERS redshift range $0.5{\le}\,z{<}1.0$.
The $\Sigma_{1}$ values can be well described as a function
of just size and mass, with little scatter in $\Sigma_{1}$ among
galaxies of a fixed size and mass. Overall, the $\Sigma_{1}$ values
increase from the top-left of the plot (large low-mass galaxies)
towards the lower-right (compact massive galaxies), as expected for a
density measure. The iso-$\Sigma_{1}$ contours notably change slope
through the plot, from being parallel to lines of constant
$\Sigma_{e}$ (magenta lines) for $\log_{10}\!\Sigma_{1\,}{\la}\,8.6$ (blue points), then
gradually becoming steeper to become parallel to lines of constant
$\sigma_{\rm inf}$ (black dashed lines) for the highest central stellar mass
densities ($\log_{10}\!\Sigma_{1\,}{\ga}\,9.2$; darker brown points). 
As a result, the upper limit of 
the blue cloud could also be defined as a threshold in $\Sigma_{1}$,
as well as $\sigma_{\rm inf}$ (or \smash{$\mathcal{M}/r_{e}$}).

Figures~\ref{smu_1kpc}b,c show the distributions of VIPERS galaxies in the
d4000--$\mathcal{M}$ plane for two redshift intervals (0.7--1.0 and
0.5--0.7), colour coded by 
$\Sigma_{1}$. In both panels, quiescent galaxies (d$4000{>}1.55$)
uniformly have high central stellar mass densities, 
$\log_{10}\!\Sigma_{1\,}{\ga}\,9.0$ (brown points). In the transition mass
regime ($10.2{<}\log_{10}\!\mathcal{M}\,{<}11.0$) where both blue cloud and red
sequence galaxies co-exist, those galaxies with low central stellar
mass densities $\log_{10}\!\Sigma_{1}{<}8.6$ (blue points) are confined within the blue cloud.
Within the blue cloud population $\Sigma_{1}$ increases linearly with
stellar mass, albeit with significant scatter, such that all massive
blue cloud galaxies with $\mathcal{M}\,{>}10^{11\,}{\rm M}_{\odot}$ also have high
central stellar mass densities $\log_{10}\!\Sigma_{1\,}{\ga}\,9.0$ (brown
points), for both redshift intervals. 
Thus, the massive blue cloud galaxy population 
($\mathcal{M}\,{>}10^{11\,}{\rm M}_{\odot}$) at $0.7{<}z{<}1.0$, many of which must
leave the blue cloud and be quenched by $z{\sim}0.5$ (Fig.~\ref{M11blue_evolution}), already have the
high $\Sigma_{1}$ values that are seemingly a prerequisite for
this quenching process to commence.

\section{Discussion}
\label{sec_discussion}

In Fig.~\ref{d4000_dist} we plotted the d4000 distributions of
galaxies as a function of mass and redshift, revealing how the
bimodal d4000 distribution seen among galaxies in the local Universe has
developed over the last eight billion years. This enabled us to track  
the evolution of the location and extent of both the blue cloud
and red sequence in the d4000--$\mathcal{M}$ plane (Fig.~\ref{d4000seq_evolution}).
This revealed how the high-mass limit of the blue cloud has retreated
significantly, from $\mathcal{M}{\sim}10^{11.2\,}{\rm M}_{\odot}$ at $z{\sim}0.9$ to
$\mathcal{M}{\sim}10^{10.7\,}{\rm M}_{\odot}$ by the present day. This retreat is what
\citet{cowie} was referring to when coining the term ``downsizing''. 

This observed downsizing of star formation can be understood by
considering the recent findings that the SFR--$\mathcal{M}$
relation of the main sequence of star-forming galaxies is not linear 
(${\rm SFR}\,{\propto}\,\mathcal{M}$) throughout as previously thought \citep{elbaz,peng10},
but shows a clear flattening at high masses \citep{whitaker,lee,tomczak16}. 
The curvature of the SFR--$\mathcal{M}$ relation increases with time,
while the stellar masses at which the sequence starts to bend
downwards decreases with time \citep{lee}.
This means that the rate of decline in the SFRs of blue-cloud galaxies is much
greater at the highest masses
($\mathcal{M}\,{\ga}10^{11\,}{\rm M}_{\odot}$), than at lower masses
($\mathcal{M}\,{\la}10^{10\,}{\rm M}_{\odot}$) where the relation is still close to linear.

We model the mass growth and sSFR evolution 
of star-forming galaxies for a range of final stellar masses (see
Appendix~\ref{SFappendix} for details),
 assuming that blue cloud galaxies assemble their stellar mass through star
formation, at rates following the evolving SFR--$\mathcal{M}$ relation
of star-forming galaxies from \citet{tomczak16}, and taking into account stellar mass loss
following \citet{moster}. 
This shows how the overall star-formation history and stellar mass
assembly (through star formation) is accelerated in the most
massive blue-cloud galaxies relative to their lower-mass
counterparts (Fig.~\ref{SFgrowth}). 
Notably, the stellar mass at which the sSFR of blue-cloud
galaxies is equal to $1/t_{0}(z)$, where $t_{0}(z)$ is the age of the
Universe at redshift $z$, is very close to the high-mass limits of the
blue cloud that we obtain over $0{<}z{<}1$. 
At $z{=}0.8$, the stellar mass of blue-cloud galaxies where ${\rm sSFR}{=}1/t_{0}$
is $\mathcal{M}{=}10^{11.10\,}{\rm M}_{\odot}$, falling to
$\mathcal{M}{=}10^{10.55\,}{\rm M}_{\odot}$ by $z{=}0$ (Fig.~\ref{SFgrowth}b). 
As d4000 can be considered a proxy for sSFR, it seems plausible that
as the sSFRs of massive blue cloud galaxies decline steadily with
time, their d4000 values correspondingly rise and eventually cross our
d$4000{=}1.55$ threshold, taking them out of our blue cloud
sample.

\subsection{The decline of massive star-forming galaxies and the rise of
  massive passive galaxies}

Figure~\ref{M11blue_evolution} showed how the number density of
massive blue-cloud galaxies ($\mathcal{M}\,{>}10^{11\,}{\rm M}_{\odot}$, d$4000{<}1.55$) has rapidly dropped five-fold between
$z{\sim}0.8$ and $z{\sim}0.5$. 
This result appears surprising given the consensus view that the SMF
of star-forming galaxies has not changed signficantly since $z{\sim}1.3$
\citep{borch,bell07,vergani08,pozzetti}. 
But these papers were based on surveys covering ${\sim}1$\,deg$^{2}$ and so strongly affected by cosmic variance. 
More recently, \citet{moustakas} 
again found no evolution over $0{<}z{<}1$ in the SMF of
star-forming galaxies at intermediate masses (10$^{9.5}$--10$^{11\,}$M$_{\odot}$) 
within the 5.5\,deg$^{2}$ PRIMUS survey. However, they also find that 
the number density of the most massive star-forming galaxies
(\smash{$\mathcal{M}\,{>}10^{11\,}{\rm M}_{\odot}$}) declines by 55\% since $z{\sim}1$.
\citet{brammer11} observe an overall ${\sim}4{\times}$ decline in the number densities of
\smash{$\mathcal{M}\,{>}10^{11\,}{\rm M}_{\odot}$} UVJ-selected star-forming galaxies between
$z{=}1.9$ and $z{=}0.6$. \citet{davidzon13} find that the most massive
($\mathcal{M}\,{>}10^{11.4\,}{\rm M}_{\odot}$) blue galaxies  
completely disappear at $z{\sim}0.6$, suggesting that at such extremely high
masses, star formation already turns off at earlier epochs ($z{>}1.3$).

Figure~\ref{M11_d4000evol} shows that these massive blue-cloud
galaxies are being steadily quenched throughout this period
($0.5{<}z{<}1.0$), transferring from the blue cloud to the red sequence.
Using the same VIPERS PDR-2 sample, \citet{gargiulo} find that the
overall number density of massive passive galaxies
\citep[\smash{$\mathcal{M}\,{>}10^{11\,}{\rm M}_{\odot}$}, NUV$rK$-selected to be
passive following][]{davidzon16}, has increased steadily from
$9.7{\pm}0.3{\times}10^{-5\,}{\rm Mpc}^{-3}$ at $z{=}0.95$ to
$16.1{\pm}0.4{\times}10^{-5\,}{\rm Mpc}^{-3}$ at $z{=}0.60$. This
increase is fully consistent with the drop in number density among the
massive blue-cloud population from $8.0{\pm}0.7{\times}10^{-5\,}{\rm
  Mpc}^{-3}$ to $1.5{\pm}0.3{\times}10^{-5\,}{\rm Mpc}^{-3}$ over the same period. 
\citet{gargiulo} show that this is also true if star-forming galaxies are selected
using the NUV$rK$ diagnostic of \citet{davidzon16}, rather than d4000 as
done here. \citet{gargiulo} split the massive passive galaxies (MPGs)
according to their mean stellar mass density $\Sigma_{r_{e}}$, and
find that while the densest MPGs ($\Sigma_{r_{e}}{>}2000\,{\rm
  M}_{\odot}\,{\rm pc}^{2}$; below the lower magenta line in Fig.~\ref{size_mass_evolution}) show essentially no evolution in
number density, the least dense MPGs ($\Sigma_{r_{e}}{<}1000\,{\rm
  M}_{\odot}\,{\rm pc}^{2}$; above the higher magenta line in
Fig.~\ref{size_mass_evolution}) show very rapid evolution in their
number density, that can explain most of the global increase in
number density of MPGs.

\subsection{The size-mass relation}

Figure~\ref{size_mass_evolution} showed that the distribution of
galaxies in the size-mass plane were confined to specific regions,
with effective radii in the range ${\sim}$1--10\,kpc, and an upper mass
limit that runs diagonally in the plots approximately along lines of constant
$\mathcal{M}/r_{e}$ or inferred velocity dispersion. These fundamental
boundaries evolve little over $0{<}z{<}1$, except that at fixed mass,
high-mass systems may be slightly more compact than seen in the local
Universe. 

In all three redshift bins, old, red sequence galaxies with
d$4000\,{\ga}1.8$ lie within a narrow, diagonal band that runs parallel
to this fundamental high-mass limit for galaxies. The dependence
of d4000 appears tilted, so that galaxies with the same d4000 value
lie along lines of constant velocity dispersion (dashed lines).
This observation parallels that seen for local
early-type galaxies within the ATLAS$^{\rm 3D}$
survey, where their $g-i$ colours, H$\beta$ absorption-line strengths,
molecular gas fractions, SSP-equivalent stellar ages, metallicities and
abundances all follow trends of constant velocity dispersion in the
size-mass plane \citep{cappellari,mcdermid}. 
\citet{shetty} find similar trends for SSP-equivalent stellar ages
following constant velocity dispersion for a sample of 154
galaxies at $0.7{<}z{<}0.9$, comprising both star-forming and quiescent galaxies.
 
\citet{whitaker16} examined the relationships between specific SFR (sSFR)
(based on stacking analysis of {\em Spitzer}/MIPS 24$\mu$m photometry) and
the location of galaxies within the size-mass plane for
$0.5{<}z{<}2.5$ galaxies from the CANDELS and 3D-HST surveys. 
They obtain very similar trends to those seen in  
Fig.~\ref{size_mass_evolution}, with specific-SFR decreasing with
stellar mass, such that galaxies along lines of constant
\smash{$\mathcal{M}/r_{e}$} have the same specific-SFRs. These trends are seen
in all three of their redshift bins: 0.5--1.0, 1.0--1.5 and
1.5--2.5. They also repeat the analysis considering only UVJ-selected star-forming
galaxies, and for their $0.5{<}z{<}1.0$ sample find the same trends as we obtain in
Figure~\ref{size_mass}, with the specific-SFRs falling for galaxies
along the leading edge of the blue cloud, and larger galaxies having
higher sSFRs than smaller ones at fixed stellar mass.
There are some hints of these trends even within their $1.5{<}z{<}2.5$
redshift bin.

Using SDSS-DR7 data, \citet{omand} examined how the fraction of
quiescent (defined using a cut in sSFR) galaxies ($f_{Q}$) depends on size
($r_{e}$) and stellar mass. For central galaxies, they find a
sharp transition from mostly active galaxies to mostly quiescent, with
contours of constant $f_{Q}$ that approximatly follow \smash{$\mathcal{M}\,{\propto}\,r_{e}^{1.5}$}. 
Among the actively star-forming galaxies, the sSFRs show a
significant dependence on effective radius at fixed stellar mass,
being $\sim$0.2\,dex lower for more compact systems, at least at
intermediate masses (\smash{$10^{10\,}{\la}\mathcal{M}\,{<}10^{11\,}{\rm M}_{\odot}$}).

\citet{lilly} caution us not to overinterpret the
trends shown in Fig.~\ref{size_mass_evolution}, or those of
\citet{omand}, 
as proof that changes in galaxy
structure must be associated with the quenching process. 
They point out the simple observation that star-forming
galaxies have undergone continual size-growth at fixed instantaneous
stellar mass of the form $r_{e\,}{\propto}\,(1+z)^{-1}$ since $z{\sim}6$ \citep{buitrago,newman,mosleh,patel,vanderwel}.  
Hence galaxies that formed their stars earlier (have larger d4000
values) will naturally be smaller than those of the same stellar mass
whose stars formed later, without any need for structure-dependent
quenching processes. Likewise, quiescent galaxies at any epoch will be
smaller than star-forming galaxies of the same stellar mass, since
the star-forming progenitors of the quiescent galaxies would have been
much smaller at the epoch at which they ceased forming stars.
\citet{lilly} consider a toy model for the growth of star-forming
galaxies in which the radial distribution of new stars being formed within
the galaxy always follows an exponential profile
with a scale length
\smash{$r_{e,SF\,}{\propto}\mathcal{M}(z)^{1/3\,}(1+z)^{-1}$}, while their
specific SFRs evolve 
to match observations of the main star-forming sequence.
This naturally ensures that ongoing star formation is always more radially
extended within these star-forming galaxies than the integrated radial
distribution of stars formed previously \citep[as observed;][]{nelson},
resulting in inside-out growth. For the most massive star-forming galaxies
(\smash{${>}\mathcal{M}^{*}_{SF}$}) central dense concentrations (``bulges'') are produced
with \smash{$\Sigma_{1}{>}10^{9\,}{\rm M}_{\odot}$\,pc$^{-2}$}, due to
the stars formed at early epochs when the galaxy was much
smaller. Steep positive radial gradients in sSFR are also
naturally produced, similar to those seen in massive star-forming
galaxies at $z{\sim}2$ \citep{tacchella}.

\citet{lilly} show that a quenching mechanism that terminates star
formation in the model galaxies with a likelihood that depends solely
on their stellar mass (and not structure), can result in 
iso-quenched fraction contours among central galaxies that lie along lines of constant surface
density $\Sigma_{r_{e}}$, closely reproducing the observations of \citet{omand}.
\citet{omand} also show that such trends can also be produced by a
quenching mechanism that depends only on \smash{$\mathcal{M}$}, but which also
reduces the galaxy size $r_{e}$ by a factor two. Finally, the sizes
and masses of already quenched galaxies can change significantly over time due to
dry mergers and repeated minor merging \citep{vandokkum,cappellari}, further altering the patterns
of $f_{Q}$ within the size-mass plane. All the above processes can
thus artificially create or alter apparent trends seen in
Figure~\ref{size_mass_evolution} or in $f_{Q}$ over the size-mass plane.

We thus argue that to determine the relevance of galaxy structure in
terminating star formation in blue-cloud galaxies, requires focusing
on the fundamental trends and boundaries within which these blue-cloud
galaxies are able to continue growing through star formation. Such an approach
should not be affected by progenitor effects or size growth through
dry/minor mergers.  
Thus the observations that the leading edge of the blue cloud in the
size-mass plane in Figs.~\ref{size_mass}--\ref{smu_1kpc} is tilted to 
run parallel to lines of constant $\mathcal{M}/r_{e}$ or
$\sigma_{\rm inf}$, rather than vertical, implies that the fundamental
limit beyond which galaxies cannot continue to form stars depends
significantly on galaxy size. Larger galaxies can continue to form stars to higher
stellar masses than smaller galaxies. 
Fig.~\ref{contours} shows how this limit retreats systematically to
lower stellar masses by 0.17\,dex from the $0.7{<}z{<}1.0$ bin to that
at $0.5{<}z{<}0.7$. This appears concrete evidence of downsizing
of star-formation at these redshifts. 
Similarly, \citet{moustakas} find that the stellar mass $\mathcal{M}_{c}$ at fixed cumulative number density
\smash{$n(\mathcal{M}{>}\mathcal{M}_{c}){=}10^{-3.5\,}{\rm Mpc}^{-3}$} of the
most massive star-forming galaxies decreases by $0.18{\pm}0.05$\,dex
since $z{=}1$, while \citet{brammer11} find a ${\sim}0.2$\,dex decrease in
$\mathcal{M}_{c}$ between $z{=}2$ and $z{=}0.5$.

As blue cloud galaxies approach the high-mass limit, their d4000 values start
to increase, pushing them to the high-d4000 wing of the blue cloud
($1.40{<}{\rm d}4000{<}1.55$). These are the galaxies most likely to
be quenched in the near future and subsequently join the red sequence. 
These quenching galaxies lie within a diagonal band (0.6--0.7\,dex wide), between the two
black dashed lines in Fig.~\ref{size_mass}a
($100{<}\sigma_{\rm inf}{<}225$\,km\,s$^{-1}$), which can be considered
the {\em quenching zone} for galaxies at these redshifts. The fact
that this quenching zone runs diagonally, combined with the 1\,dex range in
effective radii of blue-cloud galaxies, means that the range of stellar
masses where galaxies are being quenched
(0.1--3${\times}10^{11\,}{\rm M}_{\odot}$) is much wider than the width
of the quenching zone at fixed size. 

Figures~\ref{size_mass_nser} and~\ref{smu_1kpc} showed that the structures of blue-cloud
galaxies in the quenching zone have already changed, having S\'{e}rsic
indices $\eta\,{\ga}\,2$ and high central stellar mass densities
$\Sigma_{1}{>}10^{9.0\,}{\rm M}_{\odot}$\,kpc$^{2}$.
Bulge growth is necessary for these galaxies to reach the high-mass
limit of the blue cloud and be quenched. 
\citet{lang} show a marked increase in the S\'{e}rsic index and B/T ratios
of star-forming galaxies with stellar masses above $10^{11}\,{\rm
  M}_{\odot}$ at $0.5{<}z{<}2.5$ (in both redshift bins examined).
They also find that the fraction of quenched galaxies correlates strongly with bulge mass,
but not with disc mass. At fixed disc mass, the fraction of quenched
galaxies increases rapidly with B/T ratio.

It is particularly interesting that the {\em quenching zone} of
$z{\sim}0.8$ blue-cloud galaxies lies precisely along the same
size-mass relation as quiescent galaxies in the local Universe
(Fig.~\ref{size_mass_evolution}c). Given that these galaxies have
already changed their structures and have dense centres, these
quenching galaxies do not seem to have to evolve much further beyond
terminating their star formation. These objects would seem to be
likely progenitors for today's S0 population. 

 Given the well-known tight correlation between the mass of the central
supermassive black hole (M$_\bullet$) and the velocity dispersion of its host
galaxy \citep[e.g.][]{kormendyho}, the alignment of the quenching zone and the
edge of the blue cloud with lines of constant $\sigma_{\rm inf}$
is suggestive of AGN feedback playing a role in the
quenching process. \citet{vandenbosch} obtain a best-fit relation
between black hole mass M$_{\bullet}$, stellar mass and size of the
form $\log_{10}({\rm M}_{\bullet}/{\rm M}_{\odot}){=}7.48+
2.91\log_{10}(\mathcal{M}/10^{11\,}{\rm M}_{\odot})-2.77\log_{10}(r_{e}/5\,{\rm
  kpc})$, or M$_{\bullet\,}{\propto}\,(\mathcal{M}/r_{e})^{2.9}$.
As galaxies traverse the quenching zone, their central black
hole mass should thus increase by two orders of magnitude, from 
 from M$_{\bullet}{=}10^{6.28\,}{\rm M}_{\odot}$ on the
 100\,km\,s$^{-1}$ line to M$_{\bullet}{=}10^{8.33\,}{\rm M}_{\odot}$
 as they reach the 225\,km\,s$^{-1}$ line in our size-mass figures. 
\citet{heckman} show that radiative-mode AGN with accretion rates
${>}1$\% of the Eddington limit are preferentially hosted in blue-cloud
(d$4000{<}1.6$) galaxies with high stellar mass densities, exactly
describing the galaxies in our quenching zone.

Figures~\ref{smu_1kpc}b,c show that high central stellar mass
densities ($\Sigma_{1}{>}10^{9.0\,}{\rm M}_{\odot}$\,kpc$^{2}$)
appear necessary for a galaxy to have become quiescent
(d$4000{>}1.55$), as seen previously
\citep{bell08,bell12}. This threshold in $\Sigma_{1}$ also happens to
run parallel to the high-mass edge of the blue-cloud population, and
lines of constant d4000 values. This supports the idea that quenching
of star formation is linked to the presence of a dense central stellar
mass concentration in a galaxy. 
It is not that surprising that the most massive star-forming galaxies
have different structures to their lower mass counterparts. 
Using the evolving SFR--$\mathcal{M}$ relations of \citet{tomczak16} as
before to predict the star-formation histories of blue-cloud galaxies as
a function of stellar mass, the most massive blue-cloud galaxies in
VIPERS (\smash{$\mathcal{M}\,{>}10^{11\,}{\rm M}_{\odot}$}) should 
 have had their SFR peak at $z{\sim}2$ (Fig.~\ref{SFgrowth}),
before it declines by a factor 3--5 by $z{\sim}0.8$. 
Thus, the bulk of their stellar mass would have been
assembled when the galaxy was much smaller and denser, and naturally
forms a bulge-like central mass concentration (without considering any
possible compaction process). Blue cloud galaxies
which have \smash{$\mathcal{M}\,{\sim}10^{10.5\,}{\rm M}_{\odot}$} at
$z{\sim}0.8$, only had their peak SFR at $z{\sim}1$.0--1.2, and so the
bulk of star formation has occured when the galaxy is close to its
present size, meaning no significant change in
structure.

At higher redshifts, \citet{tacchella} show that massive star-forming
galaxies at $z{\sim}2.2$ (with SFRs${\sim}2$0--300\,M$_{\odot}$,
placing them on the main star-forming sequence at that redshift) already host fully grown bulges with the high
central stellar mass densities of present day quiescent ellipticals,
and show evidence of inside-out quenching. 
They suggest that the high central stellar mass densities of these
massive star-forming galaxies arise through gas-rich dissipative
processes such as violent disc instabilities leading to a compaction
of the gas disc \citep{dekel,zolotov,tacchella16}. This central mass concentration (bulge)
could then induce morphological or gravitational quenching by
stabilizing the remaining gas disc against fragmentation into
molecular clouds \citep{martig}, quenching star formation from the
inside out, resulting in central dips in the star-formation
efficiency or rings of star formation in the outer disc \citep{genzel}. 

The relative narrowness of the quenching zone and the associated
global changes in structure are indicative of a rather rigid, uniform
transformation that affects most, if not all, star-forming galaxies once they enter
the zone, and indeed before, given the rather tight SFR--$\mathcal{M}$
relation of star-forming galaxies with intrinsic scatter of 
0.2--0.3\,dex \citep{noeske,speagle,lee}. 
This uniformity would appear at odds with the merger-driven
quasar feedback model of \citet{hopkins}, whereby massive star-forming galaxies are
quenched as the result of infrequent, extreme events. This uniformity
could instead suggest that galaxies follow rather well-defined
evolutionary tracks, and that quenching is less a specific event, but a
normal continuation of their late-stage evolution \citep{gladders,abramson}.

\section{Summary}
\label{sec_summary}

We have used the full VIPERS redshift survey in combination with SDSS-DR7 
 to explore the relationships between star-formation history (as quantified using
d4000), stellar mass and galaxy structure, and establish the
limiting galaxy properties beyond which they are unable to continue
growing via star formation, and how these limits evolve from
$z{\sim}1$ to the present day. 

\begin{itemize}
\item We measured the d4000 distributions of galaxies in narrow bins of
stellar mass, in order to establish the locations and dispersions of
both the blue cloud and red sequence populations as a function of
stellar mass and redshift. We identify the upper limit of the blue
cloud as the first stellar mass bin where no appreciable peak or
feature is visible in the d4000 distribution at values indicative of
young stellar populations (d$4000{<}1.55$). The high-mass limit is
seen to retreat steadily with time from
\smash{$\mathcal{M}\,{\sim}10^{11.2\,}{\rm M}_{\odot}$} at
$z{\sim}0.9$ to \smash{$\mathcal{M}\,{\sim}10^{10.7\,}{\rm M}_{\odot}$} in galaxies from the
SDSS-DR7.

 \item The comoving number density of massive blue cloud galaxies
(\smash{$\mathcal{M}\,{>}10^{11\,}{\rm M}_{\odot}$}, d$4000{<}1.55$) is also seen to decline
by a factor 4--5 between $z{\sim}0.8$ to $z{\sim}0.5$ within the VIPERS dataset. 
These massive galaxies are being quenched in large numbers at
$z{\sim}0.7$ and contribute to the rapid increase in the number
densities of massive passive galaxies seen to occur over this period. 

\item The star-formation histories of galaxies are seen to depend on their
size ($r_{e}$) as well as stellar mass, with larger galaxies having younger
stellar populations than smaller ones at fixed stellar mass. 
Galaxies with the same d4000 values
align themselves along lines of constant \smash{$\mathcal{M}/r_{e}$} or
$\sigma_{\rm inf}$ in the size-mass plane. 

\item The upper limit of the blue
cloud population also runs along a line of constant
\smash{$\mathcal{M}/r_{e}$}. 
As star-forming galaxies grow and approach this
limit, their d4000 values start to increase, pushing them towards the
Green Valley. Their structures are also showing systematic changes,
with S\'{e}rsic indices $\eta\,{\ga}\,2.5$ more characteristic of
early-type galaxies, and high central stellar mass densities
$\Sigma_{1\,}{\ga}\,10^{9.0\,}{\rm M}_\odot$\,kpc$^{-2}$. These trends
are seen both for VIPERS galaxies at $0.5{\le}z{<}1.0$ and SDSS
galaxies in the local Universe. 

\item The massive star-forming galaxies that are leaving
the blue cloud at $z{\sim}0.8$, already lie along
size-mass relation of present day quiescent galaxies, and seemingly have to do
relatively little beyond stopping forming stars to become today's S0s. 
\end{itemize}

\begin{acknowledgements}
We acknowledge the crucial contribution of the ESO staff for the
management of service observations. In particular, we are deeply
grateful to M. Hilker for his constant help and support of this
program. Italian participation to VIPERS has been funded by INAF
through PRIN 2008, 2010, and 2014 programs. 
LG and BRG acknowledge support from the European Research Council through grant n.~291521. 
OLF acknowledges support from the European Research Council through grant n.~268107. 
JAP acknowledges support of the European Research Council through grant n.~67093. 
RT acknowledges financial support from the European Research Council through grant n.~202686. 
AP, KM, JK and MS have been supported by the National Science
Centre (grants UMO-2012/07/B/ST9/04425 and UMO-2013/09/D/ST9/04030). 
EB, FM and LM acknowledge the support from grants ASI-INAF I/023/12/0 and PRIN MIUR 2010-2011. 
LM also acknowledges financial support from PRIN INAF 2012. 
TM and SA  acknowledge financial support from the ANR Spin(e) through the french
grant  ANR-13-BS05-0005. 
SDLT and MP acknowledge the support of the OCEVU Labex (ANR-11-LABX-0060) and the A*MIDEX project
(ANR-11-IDEX-0001-02) funded by the "Investissements d'Avenir" French
government program managed by the ANR. and the Programme National
Galaxies et Cosmologie (PNCG). Research conducted within the scope of
the HECOLS International Associated Laboratory, supported in part by
the Polish NCN grant DEC-2013/08/M/ST9/00664.
\end{acknowledgements}

\begin{appendix}
\section{Stellar mass completeness limits}
\label{smcomplete}
\begin{figure}
  \centering
  \includegraphics[width=7cm]{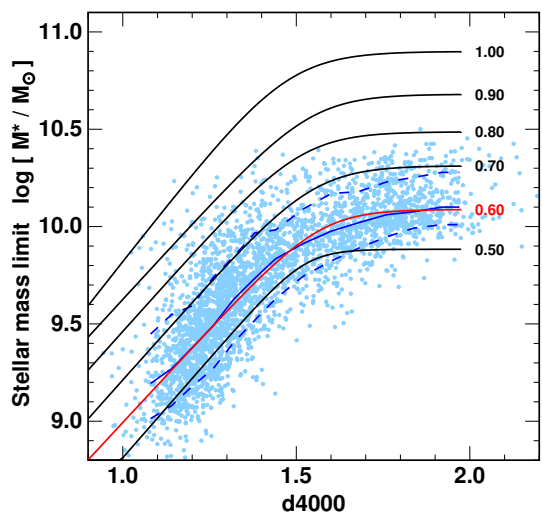}
  \caption{Stellar mass completeness limits as a function of d4000 and
    redshift. Light blue points mark the limiting stellar mass
    ($\mathcal{M}_{lim}$) of $0.575{\le}z{<}0.625$ VIPERS galaxies,
      at which they would have the apparent magnitude limit
      $i_{AB}{=}22.5$ of the survey, if placed at $z{=}0.60$. The
      solid blue curve marks the median $\mathcal{M}_{lim}$ value of
      these galaxies as a function of d4000, while the dashed lines
      mark the $1\sigma$ spread. The red curve is the best-fit double
      power-law to the blue curve. Black
      curves plot the corresponding best-fit double power-laws at
      steps of 0.1 in $z$.
  }
  \label{mass_completeness}
\end{figure}

For magnitude-limited surveys such as VIPERS, the minimum stellar mass $\mathcal{M}_{lim}$
to which a galaxy would be targetted depends on both its redshift and
its stellar mass-to-light ratio $\mathcal{M}/L$. The mass-to-light
ratio depends on the star-formation history of the galaxy, and so
should be a simple function of d4000. To derive
$\mathcal{M}_{lim}(z,d4000)$ for all redshifts, we first estimate it
for specific redshifts $z_{step}$ in intervals of 0.05 in $z$ (e.g. 0.50, 0.55,
0.60). We then take all VIPERS galaxies with redshifts within 0.025 of
$z_{step}$ and determine the $i$-band magnitude $i_{step}$ it would
have if placed at $z_{step}$. The limiting stellar mass ($\mathcal{M}_{lim}$) of that galaxy
at $z_{step}$ is then the mass it would have if its apparent magnitude
were equal to the limiting magnitude of the survey ($i_{lim}{=}22.5$)
given by $\log(\mathcal{M}_{lim})=\log(\mathcal{M})+0.4(i_{step}-i_{lim})$.
Figure~\ref{mass_completeness} plots the resulting distribution of
$\mathcal{M}_{lim}$ for galaxies placed at $z{=}0.60$ as a function of
their d4000 value (light blue points). We measure the median
$\log(\mathcal{M}_{lim})$ value of these galaxies as a function of d4000
(blue curve), and the $1\sigma$ scatter (dashed curves). The
$\mathcal{M}_{lim}$ appears to be relatively independent of d4000 for
large values (${\ga}1.6$) indicative of quiescent galaxies, but falls
exponentially with decreasing d4000 for d$4000{\la}1.4$. 
We model this by fitting the blue curve with a double power law (red
curve), that is fixed to become asympotically constant at
large d4000 values. This process is repeated for the other $z_{step}$
(black curves). These power laws represent the 50\% completeness
limits of VIPERS. By shifting each curve up and down and measuring how
the fraction of points below the curves change, the full
distribution of stellar mass completeness
$\mathcal{C}(\mathcal{M},{\rm d}4000,z_{step}$) is determined.

\section{Size-mass relation coded by the bulge-to-total ratio}
\label{BTappendix}
\begin{figure}
  \centering
  \includegraphics[width=7cm]{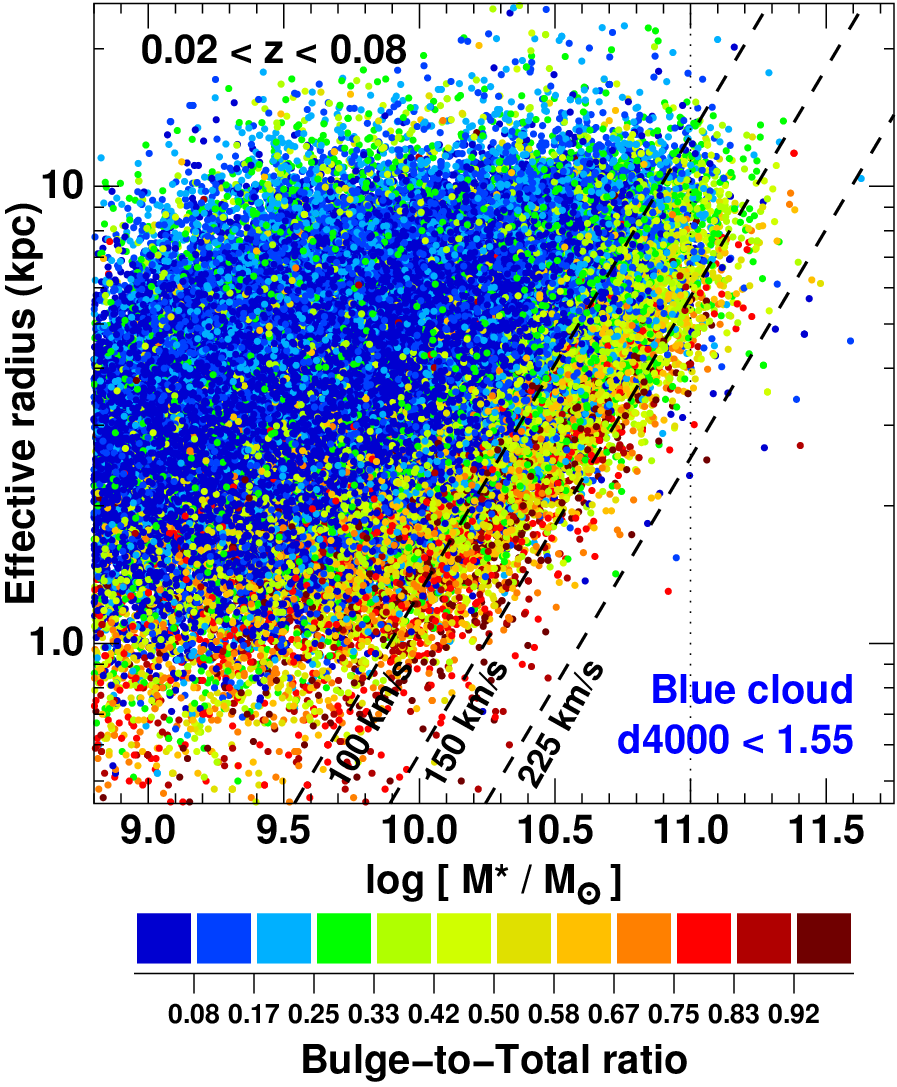}
  \caption{Size-mass relation of SDSS blue cloud galaxies
    (d$4000{<}1.55$) in the local Universe
    ($0.02{\le}z{<}0.08$). Galaxies are colour coded according to
    their bulge-to-total (B/T) ratio derived by \citet{simard}.}
  \label{size_mass_BT}
\end{figure} 

\citet{simard} performed full bulge-disc decompositions, with
$\eta{=}4$ bulge and exponential disc components, for all SDSS-DR7
galaxies in our sample. Figure~\ref{size_mass_BT} replots the
size-mass relation of blue cloud galaxies (d$4000{<}1.55$) at
$0.02{\le}z{<}0.08$ from the SDSS-DR7, colour coded according to the
fraction of $r$-band galaxian light due to the bulge component (B/T). 
For galaxies which have two distinct components, a single S\'{e}rsic
function may have difficulty fitting the radial light profile at all
radii, and so in these cases the B/T ratio should be more robust measure of galaxy structure.

Galaxies on the leading edge of the blue cloud have significant bulge components
($0.3{\la}{\rm (B/T)}{\le}1.0$), as do the most compact systems ($r_{e}{\la}1.5$\,kpc) at all
stellar masses, while the remaining low-mass galaxies are uniformly
disc-dominated (B/T${<}$0.25). This confirms that the changes in
galaxy structure reported in Fig.~\ref{size_mass_nser} based on their
S\'{e}rsic indices, for blue cloud galaxies along the high-mass edge 
of the blue cloud, do reflect concrete changes in galaxy structure.

\section{Growth and evolution of star-forming galaxies as a function
  of stellar mass}
\label{SFappendix}

\begin{figure*}
  \centering
  \includegraphics[width=18.4cm]{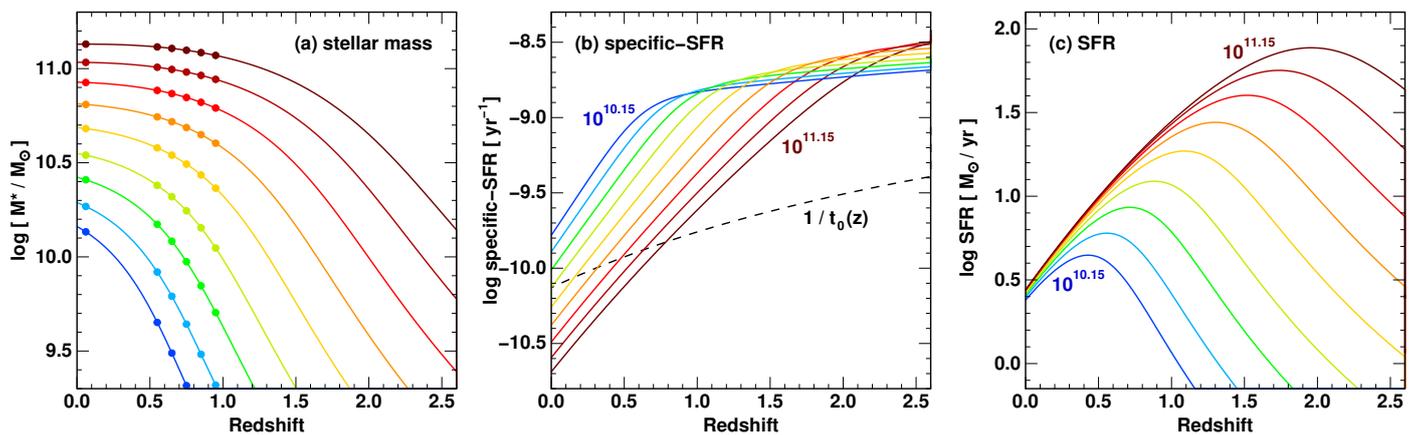}
  \caption{Mass growth profiles (left), specific-SFRs (centre) and
    SFRs (right) extracted from the evolution of the
    SFR--$\mathcal{M}$ relation of star-forming galaxies of
    \citet{tomczak16}. Mass loss due to stellar evolution is accounted
  for according to Eq.~16 of \citet{moster}. The coloured
  curves indicate how the evolution of galaxies that evolve along the 
  main star-forming sequence varies as a function of their final
  stellar mass, from low-mass galaxies ($\mathcal{M}{=}10^{10.15\,}{\rm
    M}_{\odot}$; blue) to
  those near the high-mass limit of the blue cloud ($10^{11.15\,}{\rm
    M}_{\odot}$; dark red). }
  \label{SFgrowth}
\end{figure*} 

We use the parametrized functions  describing the redshift evolution of
the SFR--$\mathcal{M}$ relation of UVJ-selected star-forming galaxies from 
\citet{tomczak} to predict the stellar mass growth of star-forming
galaxies as a function of redshift. The relations are based on
observations of $0.5{<}z{<}4$ star-forming galaxies from the FourStar
Galaxy Evolution Survey (ZFOURGE) in combination with far-infrared
imaging from {\em Spitzer} and {\em Herschel}. 
We start with a seed galaxy at
redshift 4.0, and initial stellar mass $\mathcal{M}_{0}$, from which we
calculate its predicted SFR at that redshift from the parameterized SFR--$\mathcal{M}$
relation given in Eq.~4 of \citet{tomczak}. In time steps equal to $\Delta
z{=}0.01$, the stellar mass is shifted by the amount of star
formation added according to its SFR. SFRs are recalculated at each
new time step. Mass loss due to stellar evolution is accounted for
according to Equation 16 of \citet{moster}. We then evolve the
observed stellar mass $\mathcal{M}(z)$ of the galaxy forward in time 
to the present day. We repeat this for a number of seed galaxies with
a range of initial stellar masses, in order to populate the observed stellar
mass range of star-forming galaxies at $z{=}0$. The resulting mass
growth profiles $\mathcal{M)}(z)$ are shown in Fig.~\ref{SFgrowth}a by
the coloured curves, from low-mass (blue) to high-mass (red) galaxies.

The evolutions of the specific-SFRs, SFR$(z)/\mathcal{M}(z)$, of the
same galaxies and their SFRs are shown in Figs.~\ref{SFgrowth}b and c.
The black dashed curve in Fig.~\ref{SFgrowth}b indicates the evolution
of $1/t_{0}(z)$ where $t_{0}(z)$ is the age of the Universe at redshift
$z$ in yr.
The specific-SFRs of star-forming galaxies of all masses remain
relatively constant at high redshifts at values of
${\sim}10^{-8.6}\,{\rm yr}^{-1}$ and little dependency on mass. 
Starting with the most massive star-forming galaxies they
progressively peel off this phase of flat specific-SFRs, and start a
second phase of steady decline in specific-SFR that continues unabated
to the present day. 
Massive star-forming galaxies can be seen to follow an accelerated
evolution relative to their lower mass counterparts.
Similarly, the SFRs of star-forming galaxies peak at a redshift that
increases with stellar mass (Fig.~\ref{SFgrowth}c), from $z_{\rm peak}{\sim}0.5$
for low-mass systems which have $\mathcal{M}\,{\sim}10^{10.15}\,{\rm
  M}_{\odot}$ at $z{=}0$, to $z_{\rm peak}{\sim}2.0$ for those which become today's
most massive blue-cloud galaxies with $\mathcal{M}\,{\sim}10^{11.15}\,{\rm
  M}_{\odot}$.

\end{appendix}

\end{document}